\documentclass[prc,twocolumn,preprintnumbers]{revtex4-2}
\usepackage[utf8]{inputenc}
\usepackage{amsmath}
\usepackage{amssymb}
\usepackage{physics}
\usepackage{graphicx}
\usepackage{mathrsfs,hyperref}

\usepackage{bbold}
\usepackage{xcolor}
\usepackage{comment}

\begin{document}

\title{Quantum corrections at second order in derivatives to the dynamics of small non-relativistic fluids}

\author{Lars~H.~Heyen}
\email{lars.heyen@kit.edu  }
    \affiliation{Scientific Computing Center, Karlsruhe Institute of Technology, Hermann-von-Helmholtz-Platz 1, 76344 Eggenstein-Leopoldshafen, Germany}
	\affiliation{Institut f\"{u}r Theoretische Physik, Universit\"{a}t Heidelberg, Philosophenweg 16, 69120 Heidelberg, Germany}
\author{Giuliano~Giacalone}
\email{giuliano.giacalone@cern.ch}
\affiliation{Theoretical Physics Department, CERN, CH-1211 Gen\`eve 23, Switzerland}
	\affiliation{Institut f\"{u}r Theoretische Physik, Universit\"{a}t Heidelberg, Philosophenweg 16, 69120 Heidelberg, Germany}
\author{Stefan~Floerchinger}
\email{stefan.floerchinger@uni-jena.de}
	\affiliation{Theoretisch-Physikalisches Institut, Friedrich-Schiller-Universit\"at Jena, Max-Wien-Platz 1, 07743 Jena, Germany}

\begin{abstract}
    To capture the dynamics of macroscopic non-relativistic fluids consisting of very many atoms, it is typically sufficient to truncate the gradient expansion at order zero, leading to ideal fluid dynamics, or at order one, leading to the Navier-Stokes theory. For mesoscopic fluids consisting of a small number of atoms, second-order corrections can become significant. We investigate here specifically superfluids at vanishing temperature, and identify relevant second-order terms of quantum origin that contribute already in a static situation. The general form of these terms arises from an extension of the Gross-Pitaevskii theory. In the context of density functional theory, they are named after C.~von~Weizsäcker. We assess the influence of these terms on numerical solutions of second-order fluid dynamic equations for the expansion of a mesoscopic ultra-cold Fermi gas released from an anisotropic harmonic trap in two spatial dimensions.
\end{abstract}

\preprint{CERN-TH-2024-138}

\maketitle

\section{Introduction}

We denote by \textit{mesoscopic} a system that is too large to permit a solution from first principles, and yet too small to warrant a macroscopic treatment based on conventional many-body approaches.
In recent years, the study of emergent collective phenomena in such scenarios has been the focus of much theoretical and experimental work.
A major drive for this research is the observation of collective phenomena in so-called small system collisions (such as proton-proton and proton-nucleus collisions) in high-energy experiments \cite{CMS:2010ifv,ALICE:2012eyl,ATLAS:2012cix,PHENIX:2018lia,STAR:2022pfn}. This has sparked an intense debate in regards to the formation of a quark-gluon plasma in such experiments \cite{Nagle:2018nvi,Schenke:2021mxx,Noronha:2024dtq,Grosse-Oetringhaus:2024bwr}, which would pose a challenge for all reasonable criteria of applicability of a hydrodynamic description \cite{Kurkela:2019kip,Berges:2020fwq,Ambrus:2022qya}.
Recently, emergent fluid-like phenomena, such as elliptic flow, have been observed as well in few-particle systems in the context of ultra-cold atomic experiments with controllable interactions and geometries \cite{PhaseTransition_bayha2020observing, CooperPairPaper_holten2022observation, Brandstetter:2023jsy, Lunt:2024vxs}, which calls for an improved understanding of the underlying mechanisms behind these features. In addition, much effort is also ongoing in elucidating the microscopic origin of the emergent collective properties of atomic nuclei (mesoscopic systems by definition), such as nuclear clustering or deformations, in effective field theories of low-energy QCD  \cite{Otsuka:2022bcf, Shen:2022bak, Ekstrom:2023nhc, Giacalone:2024luz, Hu:2024pee, Sun:2024iht, Frosini:2024ajq}.

In this context, a fundamental and timely question is whether fluid dynamics  as an effective dynamical framework can be applied in the mesoscopic regime. Even though hydrodynamics is developed in terms of a derivative expansion targeted at situations with a clear separation of scales \cite{Landau_Lifshitz_2013fluid}, recent results, in particular on the elliptic flow of the mesoscopic Fermi gas \cite{Brandstetter:2023jsy}, suggest that its applications could reach beyond that.
In hydrodynamic descriptions of macroscopic everyday phenomena, the derivative expansion can be truncated after the first order as gradients can be reasonably assumed to be small.
In the extreme situations we are interested in, this assumption is not guaranteed to hold.
Due to that, in this paper we focus on second-order fluid dynamics.
This has been explored extensively for relativistic fluids \cite{Jaiswal:2016hex, Romatschke:2017ejr, Rocha:2023ilf}, due to the causal behavior that a first-order truncation lacks \cite{IsraelStewart, hiscock1983stability, hiscock1985generic,  hiscock1987linear}, specifically in its application to the dynamics of the quark-gluon plasma produced in heavy-ion collisions.
On the non-relativistic side, advances have been made in the context of cold Fermi gases \cite{chao2012conformal, KIKUCHI20162075, lewis2017higher}.

We  discuss in the following the second-order contributions to both the stress-energy tensor and the heat flux in a non-relativistic context.
We provide a generic formulation of second-order fluid dynamics where the quantum pressure contributing to the Gross-Pitaevskii equation \cite{PitaevskiiStringari2016}, as well as other terms that appear via the von-Weizsäcker method in density functional theory \cite{DFT_WeizsaeckerTerm_jones1971density, DFT_parr_yang_1995, DFT_dreizler_gross_2013}, are consistently treated as second-order corrections to the gradient expansion.
As an application of this theory, we discuss mescoscopic gases of strongly interacting ultra-cold fermions in two dimensions. We calculate the influence of the aforementioned second order contributions to the Gross-Pitaevskii theory on the initial density distributions of the trapped fermions, as well as their impact on shape of the cloud during its expansion.

This paper is organized as follows. In section \ref{sec:FluidDynamics}, we recall the derivation of non-relativistic fluid dynamics in terms of conservation laws for mass, momentum and energy, together with a derivative expansion. Special emphasis is put on contributions to dissipative stresses arising at second order in derivatives. In section \ref{sec:SecondOrderCorrections}, we discuss superfluids at zero temperature. Here one can introduce a superfluid order parameter field, and make an ansatz for the quantum effective action that governs its dynamics. We discuss in particular a correction to the well-known Gross-Pitaevskii theory that can modify the coefficient of the so-called quantum pressure term, which can not be neglected for situations with small system sizes and strong interactions, and discuss how this treatment relates to density functional theory in static situations and the so-called von Weizs\"acker term. In section \ref{sec:ExpansionDynamics}, we apply our formalism to strongly-interacting ultra-cold fermions in two dimensions. We introduce an approximation of the thermodynamic equation of state, motivated by available theoretical and experimental knowledge, for the strongly interacting BEC side of the BEC-BCS crossover. Afterwards, we discuss the expansion after release from an anisotropic harmonic trap. This is first done for the well-known ideal fluid scenario, where we are able to reduce the evolution equations to ordinary differential equations that can be easily solved numerically. Subsequently, we implement second-order corrections. Numerical solutions of the resulting evolution equations are obtained both for a static system and for the expansion dynamics upon release from the harmonic trap. We comment on the comparison between our results including the second-order terms and the available data on the mesoscopic Fermi gas. Finally, we draw some conclusions in section \ref{sec:Conclusions}. Appendix \ref{app:Ambiguity} discusses some ambiguities arising for the definition of stresses at second order, while appendix \ref{app:LagrangeCoords} details how the solution to the release problem for an ideal fluid can be simplified in terms of Lagrange coordinates.

\section{Equations of Fluid Dynamics}\label{sec:FluidDynamics}

\subsection{Differential form of the conservation laws}

We recall the generic principles underlying the equations of nonrelativistic fluid dynamics, and the derivative expansion on which they are based. Similar reviews can be found in, e.g., Refs. \cite{Landau_Lifshitz_2013fluid, Schaefer:2014awa, Rocha:2023ilf}.

Non-relativistic fluid dynamics is based on on local conservation laws for particle number or mass, momentum and energy.
In their differential form, these relate the changes in densities to the divergence of corresponding fluxes.
The conservation law for mass is 
\begin{equation}
    \partial_t \rho(t,\mathbf{x})+\partial_j \rho_j(t,\mathbf{x})=0 \, ,
\end{equation}
where $\rho(t,\mathbf{x})=m n(t,\mathbf{x})$ is the mass density and $\rho_j(t,\mathbf{x})$ the corresponding current or flux density. 
The latter can be used to define the fluid velocity $v_j(t,\mathbf{x})=\rho_j(t,\mathbf{x})/\rho(t,\mathbf{x})$. As a consequence of Galilean boost symmetry, the momentum density equals the mass current, $\mathscr{P}_{k}(t,\mathbf{x}) = \rho_k(t,\mathbf{x})=\rho(t,\mathbf{x})v_k(t,\mathbf{x})$. 
The local momentum conservation law is then 
\begin{equation}
    \partial_t \mathscr{P}_k(t,\mathbf{x}) + \partial_{j}\mathscr{P}_{jk}(t,\mathbf{x})=0 \, ,    
\label{eq:localMomentumConservation}
\end{equation}
where the symmetric tensor $\mathscr{P}_{jk}(t,\mathbf{x})$ is the momentum flux density.
By its behavior with respect to Galilei transformations one can split it into several parts,
\begin{equation}
\begin{split}
    \mathscr{P}_{jk}(t,\mathbf{x}) = & \rho(t,\mathbf{x}) v_j(t,\mathbf{x}) v_k(t,\mathbf{x}) \\ & + [p(t, \mathbf{x}) + \Pi(t, \mathbf{x})] \delta_{jk} + \pi_{jk}(t,\mathbf{x}) \, ,
\end{split}\label{eq:decompositionMomentumFluxDensity}
\end{equation}
where the first term parametrizes the macroscopic fluid motion while the others transform as scalars under Galilean boosts.
Of these terms, there is a contribution to the trace, usually split into a thermodynamic pressure $p$, related to the particle number density $n$ and energy density $\varepsilon$ through the thermal equilibrium equation of state $p(n, \varepsilon)$, and a non-equilibrium bulk viscous pressure $\Pi$. Finally there is the traceless and symmetric shear stress tensor $\pi_{jk}$.

Similar to the other two conservation laws, the energy equation is formulated as
\begin{equation}
    \partial_t \mathcal{E}(t, \mathbf{x}) + \partial_j \mathcal{E}_j(t, \mathbf{x}) = 0 \, ,
\end{equation}
with the energy density $\mathcal{E}$ and the corresponding flux density $\mathcal{E}_j$.
The energy current can be split, in a similar way to the momentum flux, into 
\begin{equation}
    \mathcal{E}_j = \left( \mathcal{E} \delta_{jk} + \mathscr{P}_{jk} - \rho v_j v_k \right) v_k + q_j \, ,
\end{equation}
where the heat flux $q_j$ is introduced similarly to $\Pi$ and $\pi_{jk}$ before.
Splitting the total energy density $\mathcal{E} = \rho \mathbf{v}^2 / 2 + \varepsilon$ into a component from the macroscopic motion and the internal energy density $\varepsilon$, one finds another equivalent form of the energy equation,
\begin{equation}
    \partial_t \varepsilon + \partial_j (\varepsilon v_j + q_j) + (\mathscr{P}_{jk} - \rho v_j v_k) (\partial_j v_k) = 0 \, .
\end{equation}
Alternatively, one can formulate an evolution equation for the entropy density $s$ and entropy current. Specifically for non-dissipative fluids entropy is also conserved.

\subsection{Global thermal equilibrium}
Global thermal equilibrium states are per definition stationary, so independent of time. Moreover, they have spatially constant fluid velocity, $\mathbf{v}(t, \mathbf{x})=\mathbf{v}_0$, vanishing in the fluid rest frame, as well as constant chemical potential $\mu$ and temperature $T$. It is not necessary that the particle density $n$, the energy density $\varepsilon$ or the entropy density $s$ be constant, as well. Counterexamples are fluids in the coexistence region of a first order phase transition, or fluids in external trapping potentials.

Global thermal equilibrium states that are also spatially homogeneous are in their rest frame fully characterized by two independent thermodynamic variables such as the particle number density $n$ and internal energy density $\varepsilon$. Alternatively one could use for example temperature $T$ and chemical potential $\mu$. The pressure, entropy density, or other equilibrium properties follow from these in terms of the thermodynamic equation of state. The ``stresses'' $\Pi$, $\pi_{jk}$ and $q_j$ vanish for such homogeneous equilibrium states.

\subsection{Derivative expansion}

Besides the conservation laws, fluid dynamics is based on an expansion around spatially homogeneous thermal equilibrium states. The assumption is that the fluid state changes slowly compared to typical times needed for relaxation towards thermal equilibrium and that the fluid is spatially homogeneous enough to use the same variables and concepts as for the description of such states.

At this point one faces an ambiguity. While homogeneous global thermal equilibrium states can equivalently be described with any pair of independent thermodynamic variables, a choice has to be made for developing the gradient expansion. In particular, chemical potential $\mu$ and temperature $T$ need to be free of gradients for any global equilibrium state, while densities like $n$ and $\varepsilon$ can be non-homogeneous in the presence of non-trivial electromagnetic, gravitational or similar potentials. We see two possibilities:
\begin{itemize}
    \item[(i)] To construct a derivative expansion in terms of densities like $n$ and $\varepsilon$ where the lowest order cooresponds to homogeneous densities.
    \item[(ii)] To construct a derivative expansion in terms of conjugate variables like the chemical potential $\mu$ and temperature $T$ as well as derivatives of the external gauge potential $A_\mu =(A_0, \mathbf{A})$ like the field strength $\partial_\mu A_\nu - \partial_\nu A_\mu$.
\end{itemize}
For the present paper we follow the first option. This means to expand the stresses $\Pi$, $\pi_{jk}$, and $q_j$ in derivatives of the fluid velocity $v_j$, the particle number density $n$ and the energy density $\varepsilon$.

At lowest order is this expansion scheme, $\Pi=0$, $\pi_{jk}=0$, $q_j = 0$, one finds ideal fluid dynamics with its well-known momentum flux
\begin{equation}
    \mathscr{P}_{jk}(t,\mathbf{x}) = \rho(t,\mathbf{x}) v_j(t,\mathbf{x}) v_k(t,\mathbf{x}) + p(t, \mathbf{x}) \delta_{jk} \, .
\end{equation}
In this case the conservation equations simplify to 
\begin{equation}
\begin{split}
    [\partial_t + v_j \partial_j ] \rho + \rho \partial_j v_j & = 0 \,, \\
    \rho [\partial_t + v_j \partial_j] v_k + \partial_k p & = 0 \, , \\
    [\partial_t + v_j \partial_j ] \varepsilon + [\varepsilon + p] \partial_j v_j &= 0 \, .
\end{split}\label{eq:IdealEOM}
\end{equation}
This can be seen as a set of evolution equations for $\rho$, $\varepsilon$ and $v_k$ which gets closed by the equation of state $p(n, \varepsilon)$. Ideal fluid dynamics works reasonably well for many situations when dissipation can be neglected.

At first order in the derivative expansion scheme, bulk viscosity $\zeta$, shear viscosity $\eta$ and the heat conductivity $\kappa$, to be seen as functions of $n$ and $\varepsilon$, are introduced through 
\begin{equation}
\begin{split}
    \Pi & = - \zeta \, \partial_j v_j \, , \\
    \pi_{jk} & = - 2 \eta \sigma_{jk} = - \eta \left[ \partial_{j} v_{k} + \partial_{k} v_{j} - (2/3) \delta_{jk} \partial_l v_l \right]  \, , \\
    q_j & = -\kappa \partial_j T \, .
\end{split} \label{eq:NavierStokesStresses}
\end{equation}
In the last equation one can see the temperature as being a function of $n$ and $\varepsilon$, again expressed through the equation of state. The ansatz \eqref{eq:NavierStokesStresses} leads to the Navier-Stokes equations for the fluid velocity. 

The expressions in \eqref{eq:NavierStokesStresses} are the only corrections at first order that have to be considered because of the transformational properties with respect to Galilei boosts.
The two types of first order derivatives that are invariant under Galilean boosts are spatial derivatives and Galilean covariant derivatives $\partial_t + v_j \partial_j $, often called convective derivatives, of the fluid fields $\mathbf{v}$, $n$ and $\varepsilon$. Such covariant derivatives do not appear in the expansion above because they can be replaced to the given order by spatial derivatives using eqs.\ \eqref{eq:IdealEOM}. Also beyond first order such a replacement of the covariant derivative terms using the conservation laws is always possible. It is not exact, though, only consistent within the given order of the derivative expansion.

The only other term that could be constructed up to first order in derivatives that would transform as a symmetric tensor under rotations would be proportional to $v_j \partial_k n + v_k \partial_j n$ with some scalar coefficient function.
However, such a term cannot also transform homogeneously under Galilei boosts. Similarly, one can argue for the heat flux that the only other possible type of term (by rotation), $v_i\partial_j v_i$, is also ill-behaved with respect to Galilei boosts.

Besides Galilei and rotation symmetry, the terms in eq.\ \eqref{eq:NavierStokesStresses} are also constrained by the requirement to have a local form of the second law of thermodynamics, $\partial_t s + \partial_j s_j \geq 0$. This excludes another independent gradient term in the heat current, e.\ g. proportional to $\partial_j \mu$, and it also implies $\zeta\geq 0$, $\eta \geq 0$ and $\kappa \geq 0$.

\subsection{Superfluids}
A special case combining zeroth and first order terms in gradients is commonly used in the description of superfluids at non-zero temperature. 
In such systems an effective description is given by splitting the fluid into two parts, one with vanishing viscosity and heat conductivity (the superfluid part) and one with finite ones (the normal part), each with its own density and fluid velocity,
\begin{equation}
\begin{split}
    &\rho = \rho_n + \rho_s \, , \\
    &\rho \mathbf{v} = \rho_n \mathbf{v}_n + \rho_s \mathbf{v}_s \, .
\end{split}
\end{equation}
Thermodynamic variables such as temperature and chemical potential refer to both components. In addition to the thermodynamic information encoded in $p(T,\mu)$ one also needs a relation for the superfluid density fraction as a function of $T$ and $\mu$.

The superfluid density is given by the square of a complex superfluid order parameter, $n_s=\varphi^* \varphi$, while the superfluid velocity is the gradient of its phase, $\mathbf{v}_s =(\hbar / m) \boldsymbol{\nabla} \vartheta$, where
\begin{equation}
    \varphi(t,\mathbf{x}) = \sqrt{n_s(t,\mathbf{x})} e^{i\vartheta(t,\mathbf{x})}.
\end{equation}

As we shall further discuss in the next section, the superfluid part behaves similar to an ideal fluid, but with the additional restriction that the corresponding fluid velocity must be an irrotational potential flow. 
Any vorticity excitation in the superfluid can only exist in the form of quantized vortices (with vanishing density $n_s$ at their center).

For the normal fluid there are no such restrictions and one can make an expansion in derivatives as before.
In contrast to a one-component fluid one has to take into account that the relative velocity $\mathbf{w} = \mathbf{v}_n - \mathbf{v}_s$ is invariant under Galilean boosts and can appear accordingly.
Only the normal part is subject to dissipation and it is responsible for any heat transport and viscous damping.

Truncating after the first order in derivatives, while sufficient for most problems, can be an insufficient approximation for some. We expect this to be the case for very small fluids where gradients are necessarily large or for transition regions where the fluid density goes to zero. In such cases the effective fluid description needs to take into account higher order terms.

\subsection{Second order terms \label{sec:SecondOrderList}}

We now aim to discuss possible contributions of second order in a derivative expansion contributing to the stresses $\Pi$, $\pi_{jk}$ and $q_j$. Because all three must transform homogeneously with respect to Galilei boosts, the new terms cannot directly contain the fluid velocity but only its derivatives. Instances of the densities without any derivatives can be considered part of the coefficient function, and the only building blocks are therefore derivatives of the fluid velocity $v_j$ and densities like $n$ and $\varepsilon$.

Interestingly, while the first order corrections in eq.\ \eqref{eq:NavierStokesStresses} vanish for equilibrium configurations where fluid velocity and temperature need to be constant, this is not necessarily the case for second order corrections. Indeed they can contain derivatives of densities that might be non-vanishing for non-homogeneous thermal equilibrium states.

In order to organize the different contributions, we label them as follows,
\begin{equation}
\begin{split}
    &\Pi^{(2)} = \sum_i \Pi^{(2,i)} \, , \\
    &\pi_{jk}^{(2)} = \sum_i \pi_{jk}^{(2,i)} \, , \\
    &q^{(2)}_j = \sum_i q_j^{(2,i)} \, ,
\end{split}
\end{equation}
and use a short notation for the symmetric traceless part of a tensor,
\begin{equation}
    A_{<jk>} = \frac{1}{2} (A_{jk} + A_{kj}) - \frac{1}{D} \delta_{jk} A_{ii} \, .
\end{equation}
In this paper, we will focus on contributions to the stress-energy tensor of the types
\begin{equation}
\begin{split}
    &\Pi^{(2,1)} = \alpha_1 \partial_i \partial_i n \, , \\
    &\Pi^{(2,2)} = \alpha_2 (\partial_i n) (\partial_i n) \, , \\
    &\pi_{jk}^{(2,1)} = \beta_1 \partial_{<j} \partial_{k>} n \, , \\
    &\pi_{jk}^{(2,2)} = \beta_2 (\partial_{<j} n) (\partial_{k>} n) \, .
\end{split}\label{eq:secondOrderStressContributions}
\end{equation}
While heat conductivity, bulk viscosity, and shear viscosity are considered transport coefficients, this would not necessarily be the case for $\alpha_1$, $\alpha_2$, $\beta_1$ and $\beta_2$ as they can influence static behavior as well.
The other second order terms in the stress energy tensor can be summarized as
\begin{equation}
\begin{split}
    &\Pi^{(2,3)} = \alpha_3 (\partial_i v_i) (\partial_j v_j) \, , \\
    &\Pi^{(2,4)} = \alpha_4 (\partial_i v_j) (\partial_i v_j) \, , \\
    &\Pi^{(2,5)} = \alpha_5 (\partial_i v_j) (\partial_j v_i) \, , \\
    &\pi_{jk}^{(2,3)} = \beta_3 (\partial_i v_i) (\partial_{<j} v_{k>}) \, , \\
    &\pi_{jk}^{(2,4)} = \beta_4 (\partial_i v_{<j}) (\partial_i v_{k>}) \, , \\
    &\pi_{jk}^{(2,5)} = \beta_5 (\partial_{<j} v_i) (\partial_{k>} v_i) \, , \\
    &\pi_{jk}^{(2,6)} = \beta_6 (\partial_i v_{<j}) (\partial_{k>} v_i) \, ,
\end{split}\label{eq:secondOrderStressContributions2}
\end{equation}
which again do not contribute in equilibrium.

In relativistic contexts it is not uncommon to see one of these contributions for $\Pi$ and $\pi_{jk}$ to be replaced by an equivalent one (to the given order) containing a convective derivative,
\begin{equation}
\begin{split}
    &\Pi^{(2, \tau)} = \zeta \tau_\text{bulk} [\partial_t + v_j \partial_j] \, \boldsymbol{\nabla} \cdot \mathbf{v} \, , \\
    &\pi^{(2, \tau)}_{jk} = 2\eta \tau_\text{shear}  [\partial_t + v_j \partial_j] \sigma_{jk} 
    \, .
\end{split}
\end{equation}
The coefficients $\tau_\text{bulk}$ and $\tau_\text{shear}$ can be seen as relaxation times. 
Variants of these terms are also used e.g. in the Israel-Steward formulation of relativistic fluid dynamics \cite{Rocha:2023ilf}.

For the heat current, there are also multiple options for new contributions,
\begin{equation}
\begin{split}
    &q_{j}^{(2,1)} = \gamma_1 (\partial_j v_i) (\partial_i n) \, , \\
    &q_{j}^{(2,2)} = \gamma_2 (\partial_i v_j) (\partial_i n) \, , \\
    &q_{j}^{(2,3)} = \gamma_3 (\partial_i v_i) (\partial_j n) \, , \\
    &q_{j}^{(2,4)} = \gamma_4 \partial_i \partial_i v_j \, , \\
    &q_{j}^{(2,5)} = \gamma_5 \partial_j \partial_i v_i \, .
\end{split}
\end{equation}

The different terms presented here do not involve the energy density $\varepsilon$, but since it has the same transformational properties as the number density $n$, a generalization is straight forward. For our specific purpose in this paper such terms are not needed because we are mainly interested in fluids at vanishing temperature where energy density is non-dynamical.

A related analysis of possible second order terms appearing in the the energy momentum tensor is discussed in a relativistic context in ref.\ \cite{Romatschke:2017ejr}. 
Not counting terms containing spacetime curvature, which do not play a role for our considerations, they find the same number of contributions for both bulk viscous pressure (five terms) and shear stress tensor (six terms).

For the further analysis in this paper, we will concentrate on the second order correction terms in eq.\ \eqref{eq:secondOrderStressContributions}. We expect these terms to play a significant role also in close-to-equilibrium situations whenever spatial gradients of the density become sizeable.

We provide a complete derivation of all second-order terms  in Appendix~\ref{app:Ambiguity}, where we also discuss subtleties in the choice of these coefficients for the formulation and the solution of the second-order hydrodynamic equations.

\section{Superfluid quantum pressure as second-order term \label{sec:SecondOrderCorrections}}

Terms of second order in derivatives also arise naturally in descriptions of superfluids through their order parameter fields, and also as von-Weizsäcker type corrections in density functional theory. In both cases, the second-order terms originate from a quantum mechanical representation of the kinetic energy. 
At most places in the literature they do not appear with four independent coefficients, but only one for the von-Weizsäcker term, or none when using the Gross-Pitaevskii equation for superfluids, as we discuss now.
We focus hereafter on systems at vanishing temperature, where the particle density equals the superfluid density, $n = n_s$.

\subsection{Effective action for superfluid order parameter field}
The dynamics of a superfluid is best described in terms of an effective action for its complex order parameter field $\varphi(t, \mathbf{x}) = \langle\phi(t, \mathbf{x})\rangle$, which can be seen as (renormalized) expectation value for a suitable fundamental or composite field, $\phi(t, \mathbf{x})$.

This effective action is constrained by invariance under Galilei boosts,
\begin{equation}
    \varphi(t, \mathbf{x}) \to \exp\left(-i (m/2) \mathbf{v}^2 t / 2 + i m \mathbf{v} \cdot \mathbf{x}\right) \varphi(t, \mathbf{x} - \mathbf{v} t)  \, ,
\end{equation}
as well as a U$(1)$ symmetry (where $(A_0, \mathbf{A})$ is an external gauge field),
\begin{equation}
\begin{split}
    &\varphi(t, \vec{x}) \to e^{i \alpha(t,x)} \varphi(t, \vec{x}) \, , \\
    &A_0(t, \vec{x}) \to A_0(t, \vec{x}) + \partial_t \alpha(t, \vec{x}) \, , \\
    &\mathbf{A}(t, \vec{x}) \to \mathbf{A}(t, \vec{x}) + \boldsymbol{\nabla} \alpha(t, \vec{x}) \, .
\end{split}
\end{equation}
A covariant derivative operator adapted  to these symmetries is the combination,
\begin{equation}
    \mathscr{D} \varphi = \left[ -i\hbar (\partial_t- i A_0) - \frac{\hbar^2}{2m}(\boldsymbol{\nabla} - i \mathbf{A})^2 \right] \varphi \, ,
\end{equation}
such that the combination $\varphi^* \mathscr{D}\varphi$ transforms like a neutral scalar. Of course, also the combination $\varphi^*\varphi$ is covariant and can appear in the effective action, as well as the square of its derivative $\boldsymbol{\nabla} [\varphi^* \varphi]$.

With these elements, we can construct an ansatz for the effective action of the order parameter as a derivative expansion up to a specific order in spatial derivatives (which implicitly also limits the order in time derivatives through Galilei invariance). Up to the first non-trivial order, we write the quantum effective action as,
\begin{equation}
\begin{split}
    \Gamma = & \int \dd[d]{x} \bigg{\{} Z(\varphi^* \varphi) {\Big [} -i \hbar \varphi^* (\partial_t - i A_0) \varphi \\ & + \frac{\hbar^2}{2m}(\boldsymbol{\nabla} +i \mathbf{A}) \varphi^*(\boldsymbol{\nabla} - i \mathbf{A}) \varphi {\Big ]} + U(\varphi^* \varphi) \\ & + Y(\varphi^* \varphi) \boldsymbol{\nabla} [\varphi^* \varphi] \boldsymbol{\nabla} [\varphi^* \varphi] \bigg{\}} \, .
\end{split}\label{eq:QuantumEffectiveActionSuperfluidOrderParameter}
\end{equation}
By variation with respect to the external gauge field $(A_0$, $\mathbf{A})$, one finds the superfluid density and current:
\begin{equation}
\begin{split}
    n_s = & - \frac{1}{\hbar}\frac{\delta \Gamma}{\delta A_0} = Z(\varphi^* \varphi) \varphi^* \varphi,\\
    \mathbf{j} = & - \frac{1}{\hbar}\frac{\delta \Gamma}{\delta \mathbf{A}} = - i \frac{\hbar}{2 m} Z(\varphi^* \varphi) \left[ \varphi^* \boldsymbol{\nabla} \varphi - \varphi \boldsymbol{\nabla} \varphi^* \right].
\end{split}
\end{equation}
With the decomposition $\varphi = \sqrt{\varphi^* \varphi} e^{i\vartheta}$ the current becomes $\mathbf{j} = Z(\varphi^* \varphi) \varphi^* \varphi (\hbar / m) \boldsymbol{\nabla} \vartheta$. It is always possible, and particularly convenient, to normalize the superfluid order parameter field such that the superfluid density is $n_s=\varphi^* \varphi$. With this normalization one has effectively $Z(\varphi^* \varphi) = 1$, which we assume from now on.

\subsection{Gross-Pitaevskii equation}
Variation of the effective action \eqref{eq:QuantumEffectiveActionSuperfluidOrderParameter} with the specific choice $Y = 0$, $A_0 = - (1/\hbar)V(t,\mathbf{x})$, $\mathbf{A}=0$, with the effective potential $U(n_s)= (\lambda/2) n_s^2$ leads to the Gross-Pitaevskii equation. This describes the mean field dynamics of a system of (many) bosons with short-ranged interactions (range shorter than the typical inter-particle distance) in an external potential $V(t, \mathbf{x})$,
\begin{equation}
    i \hbar \partial_t \varphi = \left( -\frac{\hbar^2 \boldsymbol{\nabla}^2}{2m} + V + \lambda \varphi^* \varphi \right) \varphi.
\end{equation}
With the replacements
\begin{equation}
\begin{split}
    &\varphi = \sqrt{n_s} e^{i\vartheta}, \quad\quad\quad \mathbf{v} = \frac{\hbar}{m}\boldsymbol{\nabla} \vartheta, \quad\quad\quad \rho = m n_s,
\end{split} \label{eq:polarDecomposition}
\end{equation}
one obtains the Madelung-type equations,
\begin{equation}
    \partial_t \rho + \boldsymbol{\nabla} \cdot [\rho \mathbf{v}] = 0,
    \label{eq:MadelungConservationLaw}
\end{equation}
and
\begin{equation}
    \rho [\partial_t + \mathbf{v} \cdot \boldsymbol{\nabla}] \mathbf{v} + n_s \boldsymbol{\nabla} [Q + \lambda n_s + V] = 0 \, ,
\label{eq:GPMedelungRep}
\end{equation}
where $Q$ is a quantum contribution from the kinetic part of the Gross-Pitaevskii equation,
\begin{equation}
    Q = -\frac{\hbar^2}{2m} \frac{\boldsymbol{\nabla}^2 \sqrt{n_s}}{\sqrt{n_s}} \, .
\end{equation}
It is illustrating to rewrite eq.\ \eqref{eq:GPMedelungRep} in the form of a general momentum conservation law:
\begin{equation}
    \rho [\partial_t + v_j \partial_j] v_k + \partial_k [p + \Pi] + \partial_j \pi_{jk} + n_s \boldsymbol{\nabla} V(t, \mathbf{x}) = 0 \, ,
\label{eq:EulerWithStress}
\end{equation}
where $p=\lambda n_s^2/2$ is the pressure of a uniform fluid in thermal equilibrium, and (following \cite{PitaevskiiStringari2016})
\begin{equation}
\begin{split}
    \Pi = & -\frac{\hbar^2}{4 m D} n_s \boldsymbol{\nabla}^2 \ln(n_s)  
    \, , \\
    \pi_{jk} = & -\frac{\hbar^2}{4m} n_s \left[\partial_j \partial_k - \frac{1}{D}\delta_{jk} \boldsymbol{\nabla}^2 \right] \ln(n_s) \, ,
\end{split}\label{equation:stressesGP}
\end{equation}
can be understood as quantum contributions to the stresses that are of second order in gradients. Note that we have formulated eq.~\eqref{equation:stressesGP} in $D$ spatial dimensions.

This discussion shows that superfluids described by the classical Gross-Pitaevskii equation for their order parameter are still amenable to a fluid dynamical treatment based on a gradient expansion, provided that specific second order terms  of types $\Pi^{(2,1)}$, $\Pi^{(2,2)}$, $\pi_{jk}^{(2,1)}$, and $\pi_{jk}^{(2,2)}$ are included.

\subsection{Generalized superfluid equation of motion}

We now go back to the full quantum effective action in eq.\ \eqref{eq:QuantumEffectiveActionSuperfluidOrderParameter}. 
Employing again a polar decomposition as in eq.\ \eqref{eq:polarDecomposition}, with the normalization of fields $Z=1$, we can write the action, up to an irrelevant boundary term, in the alternative form
\begin{equation}
\begin{split}
    \Gamma = & \int \dd[d]{x} \bigg{\{} \hbar n_s \partial_ t \vartheta - \hbar A_0 n_s + U(n_s)  \\
    & + \frac{\hbar^2}{2m} n_s \left[ \boldsymbol{\nabla} \vartheta - \mathbf{A}\right]^2 + \left( \frac{\hbar^2}{8 m n_s} + Y(n_s) \right) [\boldsymbol{\nabla} n_s]^2 \bigg{\}} \, .
\end{split}\label{eq:QuantumEffectiveActionSuperfluidOrderParameterPolar}
\end{equation}
To find the stresses $\pi_{jk}$ and $p+\Pi$ it is best to employ Noethers theorem. Invariance of the action \eqref{eq:QuantumEffectiveActionSuperfluidOrderParameterPolar} under spatial translations yields the momentum conservation law in eq.~\eqref{eq:localMomentumConservation} with momentum flux density of the form of eq.~\eqref{eq:decompositionMomentumFluxDensity}, and
\begin{equation}
\begin{split}
\label{eq:pijk+p+Pi}
    & \pi_{jk} + [p+\Pi] \delta_{jk} \\ & = - \mathscr{L}\delta_{jk}  + \left( \frac{\hbar^2}{4 m n_s} + 2 Y(n_s) \right) \partial_j n_s \partial_k n_s.
\end{split}
\end{equation}
Here $\mathscr{L}$ is the Lagrangian density corresponding to the integrand in \eqref{eq:QuantumEffectiveActionSuperfluidOrderParameterPolar}, to be evaluated on the solution of the equation of motion. By tensor decomposition of the right hand side of eq.~\eqref{eq:pijk+p+Pi} we identify the shear stress,
\begin{equation}
\begin{split}
    \pi_{jk} = &\left( \frac{\hbar^2}{4 m n_s} + 2 Y(n_s) \right) \\
    &\times \big{[} \partial_j n_s \partial_k n_s - \delta_{jk} [\boldsymbol{\nabla} n_s]^2/D  \big{]},
\end{split}
\label{eq:shearStressYTerm}
\end{equation}
and the thermodynamic plus bulk viscous pressure
\begin{equation}
\label{eq:p+Pi}
p+\Pi = - \mathscr{L} +  \left( \frac{\hbar^2}{4 m n_s} + 2 Y(n_s) \right) [\boldsymbol{\nabla} n_s]^2/D.
\end{equation}
Then, with external gauge field $A_0= - (1/\hbar)V$, $\mathbf{A}=0$, variation of the action in eq.~\eqref{eq:QuantumEffectiveActionSuperfluidOrderParameterPolar} leads to the equation of motion
\begin{equation}
\begin{split}
    i \hbar \partial_t \varphi = & {\bigg (} - \frac{\hbar^2 \boldsymbol{\nabla}^2}{2m} + V + U^\prime(n_s) \\
    &  - 2 Y(n_s) [\boldsymbol{\nabla}^2 n_s]     
    - Y^\prime(n_s) [\boldsymbol{\nabla} n_s][ \boldsymbol{\nabla}n_s ] {\bigg )} \varphi,
\end{split}\label{eq:equationOfMotionGeneralSuperfluidOrderParameter}
\end{equation}
from which we can write the Lagrangian density in eq.~\eqref{eq:pijk+p+Pi} in the form
\begin{equation}
\begin{split}
    \mathscr{L} = & U(n_s) - n_s U^\prime(n_s) + \left[ \frac{\hbar^2}{4m} + 2 Y(n_s) n_s \right] \boldsymbol{\nabla}^2 n_s \\
    & + \left[  Y(n_s) + n_s Y^\prime(n_s) \right] [\boldsymbol{\nabla} n_s]^2.
\end{split}
\end{equation}
We can then identify the thermal equilibrium pressure for a uniform systems as:
\begin{equation}
    p = - U(n_s) + n_s U^\prime(n_s),
\end{equation}
and the bulk viscous pressure as:
\begin{equation}
\begin{split}
    \Pi = & \left[ \frac{\hbar^2}{4 m n_s D} + Y(n_s) \left(\tfrac{2}{D} - 1\right) - n_s Y^\prime(n_s)\right] [\boldsymbol{\nabla} n_s]^2 \\
    & - \left[ \frac{\hbar^2}{4m} + 2 n_s Y(n_s) \right] \boldsymbol{\nabla}^2 n_s.
\end{split} \label{eq:bulkViscousPressureYTerm}
\end{equation}
We have thus arrived at a consistent generalization of the Gross-Pitaevskii theory with more generic stress terms. 

One should note that the expressions in eqs.~\eqref{eq:shearStressYTerm} and \eqref{eq:bulkViscousPressureYTerm} do not reduce to the stresses given in eq.~\eqref{equation:stressesGP} for $Y=0$. However, the resulting fluid dynamic evolution is equivalent. This is due to the fact that the shear stress tensor and bulk viscous pressure do not directly appear in the equations, but only their derivatives or contracted products (see also Appendix~\ref{app:Ambiguity}).
In the literature this is sometimes discussed under the term ``improvement ambiguity'', which has been shown to affect the derivative expansion at second order \cite{Second_Order_Ambiguity_Nakayama}.

\subsection{Connection to density functional theory}

From eq.~\eqref{eq:QuantumEffectiveActionSuperfluidOrderParameter}, we can also write the Hamiltonian, or energy functional, as
\begin{equation}
\begin{split}
    H = & \int d^{d-1}x \left\{ 
     \frac{\hbar^2}{2m}(\boldsymbol{\nabla} +i \mathbf{A}) \varphi^*(\boldsymbol{\nabla} - i \mathbf{A}) \varphi {\Big ]} + U(\varphi^* \varphi) \right. \\ & \left. + Y(\varphi^* \varphi) \boldsymbol{\nabla} [\varphi^* \varphi] \boldsymbol{\nabla} [\varphi^* \varphi] + V \varphi^* \varphi
    \right\} \\
    =  & \int d^{d-1}x \left\{ \frac{\hbar^2}{2m} n_s \left[ \boldsymbol{\nabla} \vartheta - \mathbf{A}\right]^2 + U(n_s)  \right.  \\
    & \left. + \left( \frac{\hbar^2}{8 m n_s} + Y(n_s) \right) [\boldsymbol{\nabla} n_s]^2 + V n_s \right\}.
\end{split} \label{eq:effectiveHamiltionianWithYTerm}
\end{equation}
This can be compared to what is traditionally used in density functional theory.
A term quadratic in gradients of density also appears in the prediction of ground states using density functional theory (used in e.g. nuclear and molecular structure calculations).
In these contexts, the one-particle density is found by applying variational methods to the expectation value of the energy which has been shown to be independent of any higher order correlations (in the ground state only \cite{DFT_hohenberg_kohn_1964}).
The energy is commonly split into a kinetic and potential part plus an exchange energy that approximates the effect of interactions \cite{DFT_dreizler_gross_2013, DFT_parr_yang_1995},
\begin{equation}
\label{eq:Erho}
    E[\rho] = T[\rho] + V[\rho] + E_{xc}[\rho] \, .
\end{equation}
The terms we are looking for appear in the kinetic energy of a fermionic system \cite{DFT_weizsacker_1935, DFT_kohn_sham_1965},
\begin{equation}
    T[\rho] = \int \dd[3]{x} \left( \frac{3 (3\pi)^{2/3}}{10} \frac{\hbar^2}{m^{8/3}}  \rho^{5/3} + \lambda \frac{\hbar^2}{8m^2} \frac{(\vec{\nabla} \rho)^2}{\rho} \right) \, . 
    \label{eq:FunctionalWeizsaeckerTerm}
\end{equation}
The factor $\lambda$ is simply unity in the original work of C.~von~Weizsäcker \cite{DFT_weizsacker_1935}. In order to improve on Thomas-Fermi type predictions, later authors argued that for different systems other factors (smaller than one) are more appropriate \cite{DFT_WeizsaeckerTerm_jones1971density}.
The connection to fluid dynamics comes in when exchange terms in eq.~\eqref{eq:Erho} are neglected.
In that case, we find a differential equation for the density that is equivalent to the one obtained for hydrostatics in a system with Fermi pressure (in three dimensions) in combination with second-order terms that are also predicted by a Gross-Pitaevskii ansatz.
The main difference is the appearance of a prefactor $\lambda$ which acts as a global rescaling of the second order terms.

\subsection{Specific one parameter choice of second order terms \label{sec:SpecificYTerm}}

In the remainder of this article, we discuss numerical investigations of second-order hydrodynamics with one extra parameter. Notably, we match the effect of the von Weizsäcker term by choosing a form of the $Y$ term in the effective action of eq.~\eqref{eq:QuantumEffectiveActionSuperfluidOrderParameter} that combines with the contribution of the kinetic energy term in a way that yields an overall factor $\lambda$ in front of all second-order contributions.
We can read off directly from eqs.~\eqref{eq:shearStressYTerm} and \eqref{eq:bulkViscousPressureYTerm} that this corresponds to the relation 
\begin{equation}
    Y = \frac{\hbar^2 (\lambda - 1)}{8m n_s} \, .
\end{equation}
This is equivalent to fluid dynamics with the second order terms of eq.\ \eqref{eq:secondOrderStressContributions} replaced by:
\begin{equation}
\begin{split}
    &\Pi^{(2,1)} = \lambda \frac{\hbar^2}{8m} \boldsymbol{\nabla}^2 n_s \, , \\
    &\Pi^{(2,2)} = \lambda \frac{\hbar^2}{8m n_s} \left[\boldsymbol{\nabla} n_s\right]^2 \, , \\
    &\pi_{jk}^{(2,1)} = 0 \, , \\
    &\pi_{jk}^{(2,2)} = \lambda \frac{\hbar^2}{4m n_s} \left[ (\partial_j n)(\partial_k n_s) - \frac{1}{2} \delta_{jk} (\boldsymbol{\nabla} n_s)^2 \right] \, .
\end{split}
\label{eq:secondOrderStressLambda}
\end{equation}

Some limiting cases of interest:
\begin{itemize}
    \item $\lambda=0$ corresponds to ideal fluid dynamics;
    \item $\lambda=1$ (i.e., $Y=0$) yields the quantum pressure of Gross-Pitaevskii theory as a second-order term;
    \item $\lambda<0$ implies that the Hamiltonians in eqs.~\eqref{eq:effectiveHamiltionianWithYTerm} and \eqref{eq:FunctionalWeizsaeckerTerm} are not be bounded from below.
\end{itemize}
Note that, in this work,  all terms in eq.\ \eqref{eq:secondOrderStressLambda} involve only the superfluid density, $n_s$, which typically agrees with the full density at zero temperature. In the future, it would be interesting to extend this theory to two-component fluids at non-vanishing temperature.

\section{Impact of second-order coefficients on the  mesoscopic 2D Fermi gas}

\label{sec:ExpansionDynamics}

In this section, we apply our $\lambda$-generalized second-order hydrodynamic framework to a small-sized system of strongly-interacting ultra-cold fermions. Our motivation is mainly driven by  recent experimental breakthroughs which enable the preparation and the imaging high-fidelity pure quantum states of few fermions with tunable interactions \cite{PhysRevA.97.063613, Brandstetter:2023jsy, Subramanian_2023}. This makes our formalism amenable to potential phenomenological tests.

We start by introducing a polytropic approximation to the equation of state for the 2D Fermi gas at zero temperature which is particularly convenient for numerical and analytical implementations. We discuss then hydrodynamic solutions, both for an ideal fluid and with the inclusion of second-order corrections. We consider both a trapped (static) and an expanding system.

\subsection{Thermodynamic equation of state for the 2D Fermi gas at zero temperature}

We concentrate on a fermionic gas in a balanced mixture of two (hyperfine) spin states interacting via an s-wave contact interaction.
The strength of the latter is described in three spatial dimensions by a scattering length $a_{3D}$ that formally diverges at the unitarity point in the center of the Bardeen-Cooper-Schrieffer (BCS) to Bose-Einstein condensate (BEC) crossover \cite{PitaevskiiStringari2016, Zwerger_Bloch_Dalibard_RevModPhys.80.885, PethickSmith, Diehl:2009ma}.
In two spatial dimensions, one can define a 2D scattering length $a$ such that the binding energy of a shallow dimer is $E_B = 1/(m a^2)$.
The BCS regime corresponds to $E_B \to 0$ or $a\to\infty$ and the BEC regime to $E_B\to\infty$, $a\to 0$ with positive $a$ throughout.
In either case,
throughout the crossover one finds superfluid behavior at sufficiently low temperatures.
Moreover, in the regime around vanishing temperature, $T=0$, the normal density goes to zero, and the whole density is superfluid density.
For flow velocities below the Landau critical velocity (relative to an obstacle or boundary) superfluidity is maintained.

In the dilute regime, the thermodynamic properties depend only on a single interaction parameter, the s-wave scattering length $a$.
In order to express the thermodynamic relations in experimentally controllable quantities, it is convenient to work in the canonical ensemble, where at vanishing temperature all thermodynamic information follows from the energy density $\varepsilon(n)$ with differential $d\varepsilon = \mu dn$. By dimensional analysis is must be of the form
\begin{equation}
  \varepsilon(n) = \xi(k_\text{F} a) \varepsilon_\text{F}(n),
\end{equation}
where $k_\text{F}$ is the Fermi wave number of a free gas with two spin states and density $n$ and takes the role to parametrize the interparticle spacing. In three spatial dimensions one has $k_\text{F}(n)=(3\pi^2 n)^{1/3}$, while for two spatial dimensions one has $k_\text{F}(n)=(2\pi n)^{1/2}$. We also use the energy density of a free Fermi gas with two spin components, 
\begin{equation}
  \varepsilon_\text{F}(n) = \frac{3 k_\text{F}^2 n}{10 m} = \frac{k_\text{F}^{5/3}}{10\pi^2 m} = \frac{(243 \pi^4)^{1/3}}{10 m}n^{5/3},
  \label{eq:energyDensityFreeFermiGasD3}
\end{equation}
in three dimensions and 
\begin{equation}
  \varepsilon_\text{F}(n) = \frac{k_\text{F}^2 n}{4 m} = \frac{k_\text{F}^4}{8\pi m} = \frac{\pi}{2m} n^2,
\label{eq:energyDensityFreeFermiGasD2}
\end{equation}
in two spatial dimensions.

The function $\xi(k_\text{F} a)$ depends on the dimensionality. In three spatial dimensions one has for small negative $k_\text{F} a$ the BCS regime where by construction
\begin{equation}
  \xi(k_\text{F} a) \to 1 \quad\quad\quad (k_\text{F} a\to 0, \, k_\text{F} a<0, \text{ BCS}).
\end{equation}
The limit is approched from below because for small but non-vanishing $|k_\text{F} a|$ one has bound Cooper pairs. On the other side, for small positive $k_\text{F} a$ one has a gas of dimers of density $n/2$, with binding energy $E_\text{B}=1/(m a^2)$, and they become weakly interacting in the dilute BEC limit. There one has eventually $\varepsilon \to - n E_\text{B}/2$ and
\begin{equation}
  \xi(k_\text{F} a) \to - \frac{5}{3 (k_\text{F}a)^2}  \quad\quad\quad (k_\text{F} a\to 0, \, k_\text{F} a>0, \text{ BEC}).
\end{equation}
This limit is approached from above because the dimers have a repulsive interaction.

At the unitarity point one defines
\begin{equation}
  \xi(k_\text{F} a) \to \xi_\text{B} - \frac{5\pi \alpha}{2 (k_\text{F}a)} \quad\quad\quad (k_\text{F}a\to \pm \infty, \text{ Unitarity}),
\end{equation}
where $\xi_\text{B} \approx 0.4$ is the Bertsch parameter \cite{Bertsch1, Bertsch2}, and $\alpha\approx 0.12$ relates the Tan contact at unitarity to the Fermi wavenumber (determined numerically in e.g. \cite{Carlson_PhysRevA.83.041601, Forbes_PhysRevA.86.053603}).

In two spatial dimensions $k_\text{F} a$ is positive everywhere and the BCS limit corresponds to vanishing binding energy, $E_\text{B}=0$, or large $k_\text{F} a$,
\begin{equation}
  \xi(k_\text{F} a) \to 1 \quad\quad\quad (k_\text{F} a\to \infty, \text{ BCS}).
\end{equation}
In contrast, large binding energies correspond to the BEC limit where eventually $\varepsilon \to - n E_\text{B}/2$ or
\begin{equation}
  \xi(k_\text{F} a) \to - \frac{2}{(k_\text{F}a)^2} \quad\quad\quad (k_\text{F} a\to 0, \text{ BEC}).
\end{equation}
It is possible to supplement the different limits by next-to-leading order corrections for weakly interacting fermions, or bosonic dimers, or in the vicinity of the unitary regime in terms of the Tan contact \cite{PitaevskiiStringari2016}. However, this is beyond our purpose.

We observe that in all limiting cases discussed above the internal energy density is related to the particle density by a power law,
\begin{equation}
  \varepsilon(n) = \text{const} \times n^\kappa.
  \label{eq:powerLawEOS}
\end{equation}
In the BEC limit of free dimers, $\kappa=1$ and the constant prefactor is given by $-E_\text{B}/2$. In the BCS limit and in the unitarity regime in three dimensions, one has $\kappa = 5/3$ and the constant can be read off from \eqref{eq:energyDensityFreeFermiGasD3} with an additional factor $\xi_\text{B}$ at the unitarity point. This exponent follows from scale invariance for fermions with vanishing or infinite scattering length. In contrast, in two spatial dimensions the BCS limit corresponds to $\kappa=2$ as also implied by scale invariance. In intermediate regimes one cannot expect a similar power law, however, as it is not possible to connect all physical situation so simply. On the other hand, a modified energy density with the trivial binding energy $\varepsilon_\text{B} = -n E_\text{B}/2$ subtracted may be to very good approximation given by a power law of the density throughout the crossover (see appendix \ref{app:LagrangeCoords} for a proposal). Consequently, a similar relation holds for the pressure,
\begin{equation}
    p(n) = g n^\kappa \, . \label{eq:polytropicPresEOS}
\end{equation}
The equation of state \eqref{eq:polytropicPresEOS} is a simplified ansatz which we expect to be reasonable in two dimensions where correlations have a substantial impact also for an interacting Bose gas.
This is the case we will focus on.

Beyond these considerations, the equation of state Ansatz in eq.~\eqref{eq:polytropicPresEOS} is strongly supported by numerical \cite{PhysRevLett.106.110403} and experimental studies \cite{Jochim_Enss_PhysRevLett.116.045303, Turlapov_PhysRevLett.112.045301}.
These works focus on two-dimensional Fermi gases specifically, where the crossover is parametrized in terms of the logarithm $\eta = \ln(k_F a)$, mapping the BCS limit to $\eta\to\infty$ and the BEC limit to $\eta\to-\infty$. The system under study is a macroscopic gas of $^{6}$Li atoms, whose interactions can be nicely tuned thanks to a broad Feshbach resonance.
Our applications focus on the same experimental setup of the recent experimental work in ref.~\cite{Brandstetter:2023jsy}, where the relevant interaction strength range is $-1.7 < \eta < 1.2$.
We fit in particular an equation of state of the form \eqref{eq:polytropicPresEOS}, or $p/p_F = \alpha_1 e^{\alpha_2 \eta}$ to the experimental data of ref. \cite{Turlapov_PhysRevLett.112.045301}.
The best fit was found for
\begin{equation}
    \alpha_1 \approx 0.216 \, , ~ ~ \alpha_2 \approx 0.67 \, .
\end{equation}
One can also write this as
\begin{equation}
    p_\text{fit}(n) = \alpha_1 \left( 2\pi a^2 \right)^{\alpha_2 / 2}\frac{\pi}{2m} n^{2+\alpha_2 / 2} \, ,
\end{equation}
where we recall that $a$ is the two-dimensional $s$-wave scattering length, which is determined by the experimental conditions \cite{Brandstetter:2023jsy}, while $m=6$ a.u. is the mass of the $^{6}$Li atom.

\subsection{Scaling flow solution for ideal fluid \label{sec:idealScalingFlow}}

With the equation of state in hand, we have now closed the system of equations that governs the movement of a fluid up to zeroth order in the derivative expansion (ideal fluid).
For strongly interacting fermionic atoms at almost zero temperature one also expects superfluidity.
This excludes any viscous corrections.
Second order terms as discussed in sections \ref{sec:SecondOrderList} and \ref{sec:SecondOrderCorrections} are expected to be subleading when the fluid fields are smooth as is usually the case for macroscopic fluids.
Under these conditions we can now solve the equations of motion for a two-dimensional cloud of atoms released from an anisotropic harmonic trapping potential. We shall now study this ideal fluid flow problem before taking second order terms into account in a second step.

\subsubsection{Hydrostatic problem}

The first part of this problem is the hydrostatic one.
While the trap is still active, the fluid is assumed to be in equilibrium, i.e. its velocity vanishes everywhere and the density is constant in time.
The continuity equation is trivially fulfilled, but the momentum equation gives the condition
\begin{equation}
    \nabla p + n \nabla V = 0 \, .
\end{equation}
Using $dp=sdT + nd\mu$ (where $dT=0$ in our considerations) this can be rewritten in the typical form of a Thomas-Fermi type condition for an equilibrium density in a trap (wherever the density does not vanish),
\begin{equation}
    \mu(n) = \mu_0 - V = \mu_0 - \frac{m}{2} \omega^2_{jk} x_j x_k \, , 
\label{eq:ThomasFermiChemicalPotential}
\end{equation}
where $\mu_0$ is an integration constant that fixes the total particle number and the symmetric frequency matrix $\omega^2_{jk}$ can be taken to be diagonal by choice of a convenient coordinate system, $\omega^2_{jk} = \omega_j^2 \delta_{jk}$.
For a polytropic equation of state as in eq.~\eqref{eq:polytropicPresEOS} with $\kappa > 1$, the chemical potential can also be expressed as a power of the density,
\begin{equation}
    \mu(n) = \frac{\kappa g}{\kappa - 1} n^{\kappa - 1} \, ,
\end{equation}
which leads to the typical Thomas-Fermi density profile
\begin{equation}
    n(\mathbf{x}) = \left( \frac{(\kappa-1)(\mu_0 - \frac{m}{2} \omega^2_{jk} x_j x_k)}{g \kappa} \right)^{\frac{1}{\kappa - 1}} \, .
    \label{eq:TFDensityProfile}
\end{equation}
This formula is valid within the area defined by $\mu_0 > \frac{m}{2} \omega^2_{jk} x_j x_k$, outside of that the density is zero. Note that the value of $\mu_0$ is fixed by the normalization of the particle density, i.e., by the total mass.

It is interesting to consider moments of coordinates of this density distribution in the harmonic trap,
\begin{equation}
  \langle x_j x_k \rangle = \frac{\int d^D x \; x_j x_k \, n(\mu_0 - V(x))}{\int d^D x \; n(\mu_0 - V(x))}. 
  \label{eq:MomentsInitialDensity}
\end{equation}
Performing the variable substitution $z_j = \omega_j x_j$ one can reduce this to an isotropic integral
\begin{equation}
  \langle x_j x_k \rangle \omega_j \omega_k =\frac{\int dz \, z^{D+1} \, n(\mu_0 - mz^2 / 2)}{\int dz \, z^{D-1} \, n(\mu_0 - mz^2 / 2)}.
\end{equation}
This works similarly for higher order moments or when particle density is replaced by entropy density, energy density or similar. In particular, ratios of such moments are universally given by
\begin{equation}
  \langle x_j^2 \rangle / \langle x_k^2 \rangle = \omega_k^2 / \omega_j^2. 
  \label{eq:ratioxjxkInitial}
\end{equation}
This is independent of the thermodynamic equation of state and a consequence of the fluid statics in the form of the Thomas-Fermi approximation only. 

\subsubsection{Evolution equations}

We turn now to the dynamics. For both superfluids and normal fluids, the problem of the expansion of a fermi gas from an anisotropic trap has been discussed at length in the literature \cite{Schafer:2009zjw}, see also ref.~\cite{Li:2024ivj} for a new experimental investigation. In particular, when the aspect ratio of the trapped gas is close to zero, analytical solutions leading to simple scaling flow solutions can be obtained, as originally pointed out in ref.~\cite{PhysRevLett.89.250402}.  Here we relax this condition, and perform full numerical calculations with arbitrary trap frequencies. In addition, we point out a solution scheme based on Lagrangian coordinates which largely simplifies the generic problem.

We are interested in is the expansion of the density profile given by the form of eq.~\eqref{eq:TFDensityProfile} into free space after the confining trap is quenched off instantaneously.
Specifically for the Thomas-Fermi initial density profile it is possible to simplify the problem such that one only needs to solve for a time (but not space) dependent scaling matrix $J_{jk}(t)$ (and its inverse $I_{kj}$), see e.\ g.\ ref.~\cite{PitaevskiiStringari2016}.
This reformulation can be conveniently done by using Lagrangian coordinates (see appendix \ref{app:LagrangeCoords}).
The evolution of density and fluid velocity is then completely fixed as
\begin{equation}
\begin{split}
    &n(t,x) = \frac{n(t_0, I_{kj}(t) x_j)}{\det(J(t))} \, , \\
    &v_k(t,x) = I_{ij}(t) \dot{J}_{jk}(t) x_i \, .
\end{split}
\end{equation}
This automatically fulfills the continuity equation and the momentum equations reduces to the ordinary differential equation
\begin{equation}
    (\det(J))^{\kappa - 1} J_{jk} \ddot{J}_{lk} = \omega^2_{jl} \, ,
    \label{eq:eomJ}
\end{equation}
with the initial condition
\begin{equation}
    J_{jk}(t_0) = \delta_{jk} \, , \quad\quad\quad \dot{J}_{jk}(t_0) = 0 \, .
\end{equation}
If the coordinate system is chosen such that the frequency matrix $\omega^2_{jk}$ is diagonal, so will be the rescaling matrix $J_{jk}$.

For an isotropic trap profile $\omega^2_{jk} = \omega^2 \delta_{jk}$ the Jacobi matrix is also isotropic $J_{jk}(t)= \zeta(t) \delta_{jk}$ and the equation of motion \eqref{eq:eomJ} becomes
\begin{equation}
  \zeta(t)^{1+(\kappa-1)D}\frac{d^2}{dt^2} \zeta(t) = \omega^2,
  \label{eq:eomIsotropic}
\end{equation}
with initial conditions $\zeta(t_0)=1$ and $\frac{d}{dt}\zeta(t_0) = 0$.

Scale invariance corresponds to $\kappa = 1+2/D$ and in that case the solution is easily found to be
\begin{equation}
  \zeta(t) = \sqrt{1+\omega^2 (t-t_0)^2} \, .
\end{equation}
More generally one can find an implicit solution in the form
\begin{equation}
\begin{split}
  &_2F_1\!\left( \tfrac{1}{2}, -\tfrac{1}{(\kappa-1)D}; 1-\tfrac{1}{(\kappa-1)D}; \zeta(t)^{-(\kappa-1)D} \right) \zeta(t) = \\
  &\sqrt{\tfrac{2}{(\kappa-1)D}} \omega(t-t_0) + c_F,    
\end{split}
\end{equation}
where ${}_2F_1$ is a hypergeometric function and 
\begin{equation}
\begin{split}
    c_F &= {}_2F_1\!\left( \tfrac{1}{2}, -\tfrac{1}{(\kappa-1)D}; 1-\tfrac{1}{(\kappa-1)D}; 1 \right) \\
    &= \frac{\sqrt{\pi}\Gamma(1-\frac{1}{(\kappa-1)D})}{\Gamma(\frac{1}{2}-\frac{1}{(\kappa-1)D})},    
\end{split}
\end{equation}
its value for $\zeta(t) = 1$.

While it is hard to get from this to an explicit form for $\zeta(t)$, we can already draw some conclusions about its asymptotic behavior. Considering that the second time derivate $\ddot \zeta(t)$ is strictly positive as long as $\zeta(t)$ itself is strictly positive according to equation \eqref{eq:eomIsotropic}, the initial conditions imply that $\zeta(t)$ is strictly monotonously increasing in time. As such, the last argument of the hypergeometric function will vanish for $\kappa > 1$ at $t\to\infty$.
We make use of the identity $_2F_1(a,b;c;0) = 1$ to find the asymptotic behavior
\begin{equation}
    \lim_{t\to\infty} \frac{\zeta(t)}{t} = \sqrt{\tfrac{2}{(\kappa-1)D}} \omega \, .
\end{equation}
We find asymptotically linear scaling which we associate with free streaming at late times.

Beyond the isotropic case one can solve \eqref{eq:eomJ} numerically, or possibly in a perturbative scheme for small anisotropy. Working in the frame where the trap frequencies are diagonal, $\omega^2_{jk} = \omega^2_k \delta_{jk}$ and similarly the Jacobi matrix, $J_{jk}(t) = \zeta_k(t) \delta_{jk}$ the equations of motion are
\begin{equation}
  \det(J(t))^{\kappa-1} \zeta_{k}(t) \ddot \zeta_{k}(t) = \omega^2_{k},
  \label{eq:reducedJEom}
\end{equation}
where $k=1,\ldots, D$ and $\det(J(t)) = \zeta_1(t) \cdots \zeta_D(t)$. 
Assuming that both the frequency matrix are only slightly perturbed with respect to the isotropic solution,
\begin{equation}
\begin{split}
    \zeta_k = \zeta + \delta \zeta_k \, , \\
    \omega_k^2 = \omega^2 + \delta \omega^2_k \, ,
\end{split}
\end{equation}
and after linearising in the perturbations, we find the equation of motion
\begin{equation}
    \zeta^{1+D(\kappa-1)} \delta \ddot{\zeta}_k + \frac{\omega^2}{\zeta} \left( \delta \zeta_k + D(\kappa - 1) \sum_{j=1}^D \delta \zeta_j \right) = \delta \omega^2_k \, ,
\end{equation}
with the initial conditions $\delta \zeta_k(t_0) = 0$ and $\delta \dot{\zeta}_k(t_0) = 0$.
Since the isotropic case is solved separately, we can without loss of generality assume the perturbation of the frequency matrix is traceless, i.e.
\begin{equation}
    \sum_{j=1}^D \delta \omega^2_j = 0 \, .
\end{equation}
With the given initial conditions, this automatically implies that a similar relation also holds for the perturbation in the scaling functions,
\begin{equation}
    \sum_{j=1}^D \delta \zeta_j = 0 \, .
\end{equation}
With this, the equations of motion decouple and we obtain
\begin{equation}
    \delta \ddot{\zeta}_k + \omega^2 \zeta^{-(1+D(\kappa-1))} \left( \frac{\delta \zeta_k}{\zeta} - \frac{\delta \omega^2_k}{\omega^2} \right) = 0 \, .
\end{equation}
This is an inhomogeneous second order ordinary differential equation for which one can quickly find the particular solution
\begin{equation}
    \frac{\delta \zeta^\text{(part)}_k}{\zeta} = \frac{\delta \omega^2_k}{2\omega^2} \, ,
\end{equation}
which solves the differential equation, but does not vanish at initial time.
For the special case $\kappa = 1 + 2/D$, we find an explicit solution for the given initial conditions,
\begin{equation}
    \delta \zeta_k (t) = \frac{1}{2} \frac{\delta \omega^2_k}{\omega^2} \sqrt{1 + \omega^2 t^2} \left[1 - \cos(\sqrt{2} \tan^{-1}(\omega t)) \right] \, .
\end{equation}

Let us consider moments of the density distribution similar to \eqref{eq:MomentsInitialDensity}, but now at time $t$,
\begin{equation}
  \langle x_j x_k  \rangle(t) = \frac{\int d^D x \; x_j x_k \, \rho(t, \mathbf{x})}{\int d^Dx \; \rho(t, \mathbf{x})},
\end{equation}
where $\rho$ is the mass density. In Lagrangian coordinates this is easily evaluated to be
\begin{equation}
  \langle x_j x_k  \rangle(t) =  J_{mj}(t) J_{nk}(t) \langle x_m x_n \rangle(t_0).  
\end{equation}
In the frame where the the trap frequency tensor is diagonal this yields for ratios
\begin{equation}
\frac{\langle x_j^2 \rangle(t)}{\langle x_k^2 \rangle(t)} = \frac{\zeta_j(t)^2}{\zeta_k(t)^2} \frac{\langle x_j^2 \rangle(t_0)}{\langle x_k^2 \rangle(t_0)} =  \frac{\zeta_j(t)^2}{\zeta_k(t)^2} \frac{\omega_k^2}{\omega_j^2}
\end{equation}
In the last step we used eq.\ \eqref{eq:ratioxjxkInitial}. As a response to initial pressure gradients one expects that such ratios of moments approach first unity before they get inverted at later times. This is in contrast to the expansion dynamics of an ideal gas (ballistic expansion) where one expects an isotropic aspect ratio at asymptotically large times. 

\subsubsection{Application to the mesoscopic 2D Fermi gas}

We evince that the solutions to this problem are fully determined by the two trap frequencies, $\omega_x$ and $\omega_y$, and the two parameters $g$ and $\kappa$ of the polytropic equation of state in eq. \eqref{eq:polytropicPresEOS}. The solution does not depend on the particle number or total mass in the system. Reference~\cite{Brandstetter:2023jsy} indicates a many-body limit for the aspect ratio at asymptotically large expansion times of
\begin{equation}
\label{eq:drxdry}
  \sqrt {  \frac{\langle x_j^2 \rangle(t\rightarrow \infty)}{\langle x_k^2 \rangle(t\rightarrow \infty)} } 
  =  0.52,
\end{equation}
which is obtained by performing the ideal hydrodynamic expansion of a Thomas-Fermi initial profile with the same polytropic equation of state discussed here. There, it is found that the result does not depend on the total mass or atom number, $N_{\rm atom}$. Therefore, the present derivation provides a clear explanation for this behavior. 
\begin{figure*}[t]
    \centering
    \includegraphics[width=\textwidth]{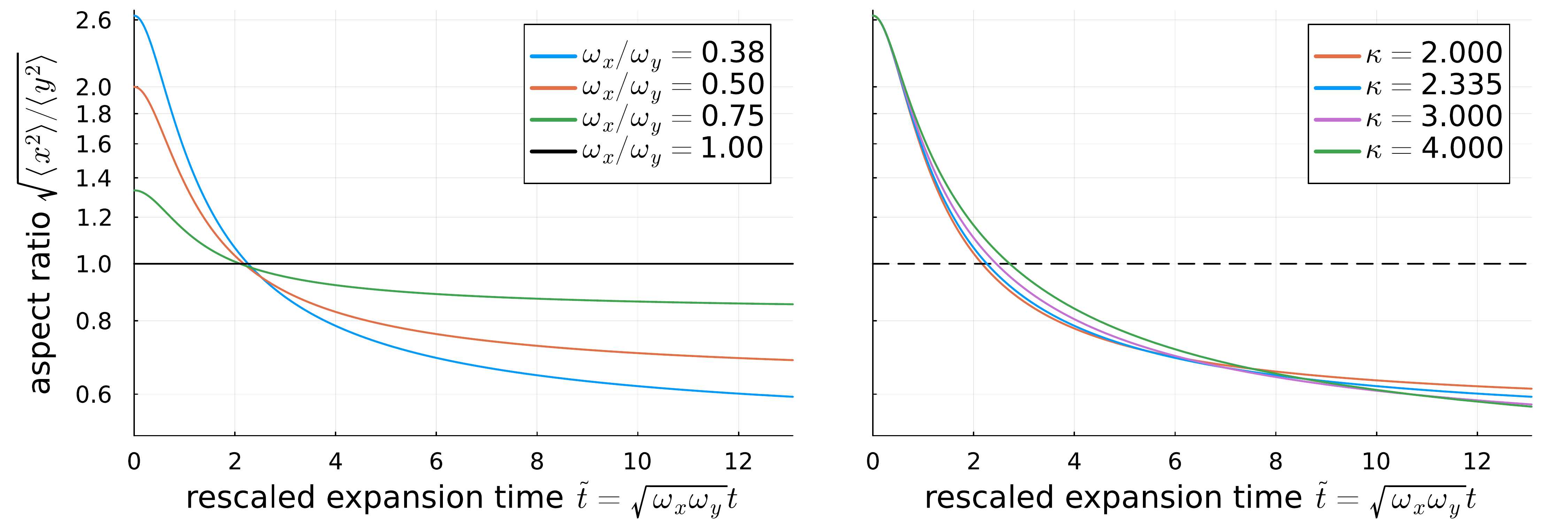}
    \caption{
        (left) Time evolution of the aspect ratio of the ideal fluid with Thomas-Fermi initial conditions for different trap frequency ratios at fixed geometric average;
        the polytropic exponent is $\kappa=2.335$ as defined in eq. \eqref{eq:polytropicPresEOS}; 
        the blue curve corresponds to the experimentally used frequencies in \cite{Brandstetter:2023jsy} 
        (right) time evolution of the aspect ratio of the ideal fluid with Thomas-Fermi initial conditions for different polytropic exponents as defined in eq. \eqref{eq:polytropicPresEOS}; the trap frequency ratio is fixed at $\omega_x / \omega_y = 0.38$ such that the blue curves match between both diagrams.
    \label{fig:TF_aspectratios_over_time}
    }   
\end{figure*}

We now take a look at some numerical solutions of the scaling flow problem for the ideal fluid in two dimensions, by solving numerically the equations discussed above. The dynamics of the fluid is described by
\begin{equation}
\begin{split}
    &\rho(t, \mathbf{x}) = \frac{\rho(t = 0, (x / \zeta_x(t), y /\zeta_y(t))^T)}{\zeta_x(t) \zeta_y(t)} \, , \\
    &v_j = \frac{\dot{\zeta}_j(t)}{\zeta_j(t)} x_j \, , \quad\quad\quad j \in \{x, y\} \, , \\
    &\ddot{\zeta}_x(t) - \omega_x^2 \zeta_x(t)^{-\kappa} \zeta_y(t)^{1 - \kappa} = 0 \, , \\
    &\ddot{\zeta}_y(t) - \omega_y^2 \zeta_y(t)^{-\kappa} \zeta_x(t)^{1 - \kappa} = 0 \, ,
\end{split}
\label{eq:scalingHydroEquations}
\end{equation}
with the initial conditions $\zeta_x(t=0) = \zeta_y(t=0) = 1$, $\dot{\zeta}_x(t=0) = \dot{\zeta}_y(t=0) = 0$ (see also eq. \eqref{eq:reducedJEom}).
We evaluate the aspect ratio $\sqrt{\expval{x^2} / \expval{y^2}}$ as a function of time.
We solve eqs.~\eqref{eq:scalingHydroEquations} with a fourth order accurate Rosenbrock method (Rodas4 solver in the Julia module DifferentialEquations). The results are displayed in Fig.~\ref{fig:TF_aspectratios_over_time}. The final aspect ratio of the cloud appears to be determined by the initial one (or equivalently by the ratio of the trap frequencies). On the other hand, the time at which the aspect ratio becomes unity is only weakly dependent on the initial aspect ratio for a fixed polytropic exponent (left panel of the figure).
Increasing the polytropic exponent (right panel of the figure) has some impact on the crossing time, and changes only slightly the large-time aspect ratio.
Notably, all these results are independent of the prefactor of the polytropic equation of state which does not appear in the differential equations for the scaling functions.
Any dependence on the geometric average of the trap frequencies can be absorbed into a rescaling to a dimensionless time variable $\tilde{t} = \sqrt{\omega_x \omega_y} t$.
In this variable, the aspect ratio becomes unity around $\tilde{t} \approx 2.1$ $\mu$s in most of the cases we considered. Note that, when matching all parameters to the experimental setup of ref.~\cite{Brandstetter:2023jsy}, we recover the numerical value $\sqrt{\langle x^2\rangle} / \sqrt{\langle y^2 \rangle}  = 0.52$ for the asymptotic aspect ratio in eq.~\eqref{eq:drxdry}.

\subsection{Second order corrections to hydrostatics \label{sec:hydrostatics}}

As mentioned before, when dealing with small system sizes or low particle numbers, quantum corrections to ideal fluid dynamics encoded by second-order terms may arise, even in the superfluid scenario. In particular, the Thomas-Fermi density is not smooth at the boundary of the atom cloud, and this issue only gets worse when the system is small. It is perhaps not surprising, thus, that a zeroth-order or ideal-fluid truncation does indeed not seem to capture some of the experimental observations made for the initial condition and the expansion 
of a cloud of 10 atoms \cite{Brandstetter:2023jsy}. We discuss first the corrections to the hydrostatic problem arising from second-order terms.

As mentioned in section \ref{sec:SecondOrderList}, the equilibrium density profile at vanishing temperature cannot be modified by first order corrections, but only by second order ones of types $\Pi^{(2,1)}$, $\Pi^{(2,2)}$, $\pi_{jk}^{(2,1)}$, and $\pi_{jk}^{(2,2)}$, as defined in eq.~\eqref{eq:secondOrderStressContributions}.
For the Thomas-Fermi equilibrium solution of a trapped gas with a polytropic equation of state, we can analytically calculate the density profile which leads to the result in eq. \eqref{eq:TFDensityProfile}.
If we want to include the second order terms presented in section \ref{sec:SpecificYTerm}, the equation we need to solve becomes, with $\boldsymbol{\nabla} p = \kappa g \rho^{\kappa - 1} \boldsymbol{\nabla} \rho$,
\begin{equation}
    -\lambda \frac{\hbar^2}{2m^2} \rho \boldsymbol{\nabla} \left(\frac{\boldsymbol{\nabla}^2 \sqrt{\rho}}{\sqrt{\rho}} \right) + \boldsymbol{\nabla} p + \frac{1}{m}\rho \boldsymbol{\nabla} V = 0.
\label{eq:initialDensityEquationWithYTerm}
\end{equation}
Solving this for the density is equivalent to finding a real-valued ground state for the Hamiltonian in eq.~\eqref{eq:effectiveHamiltionianWithYTerm}.
We do this by employing the split step Fourier method.
The time evolution operator $U(\delta t) = \exp(i \, H \, \delta t)$ is split into two exponentials, respectively, a kinetic and an interaction term, and neglecting contributions of order $\delta t^2$.
The kinetic part is applied in a momentum space representation of the order parameter, the interaction part in a real space one.
For $\lambda > 0$ the latter contains a term proportional to $(\boldsymbol{\nabla} n)^2$ which is calculated in a central finite difference scheme.
An initially Gaussian profile is then evolved with the so approximated time evolution operator in imaginary time, with the normalization being fixed by the desired number of particles at every time step.
Through this imaginary time evolution the state evolves towards the ground state as all higher energy states decay faster in comparison.
The initial density profile, i.e., the approximate solution to eq.~\eqref{eq:initialDensityEquationWithYTerm}, is then obtained from the order parameter as $\rho=m \varphi^* \varphi$. 
\begin{figure*}
    \centering
    \includegraphics[width=0.80\textwidth]{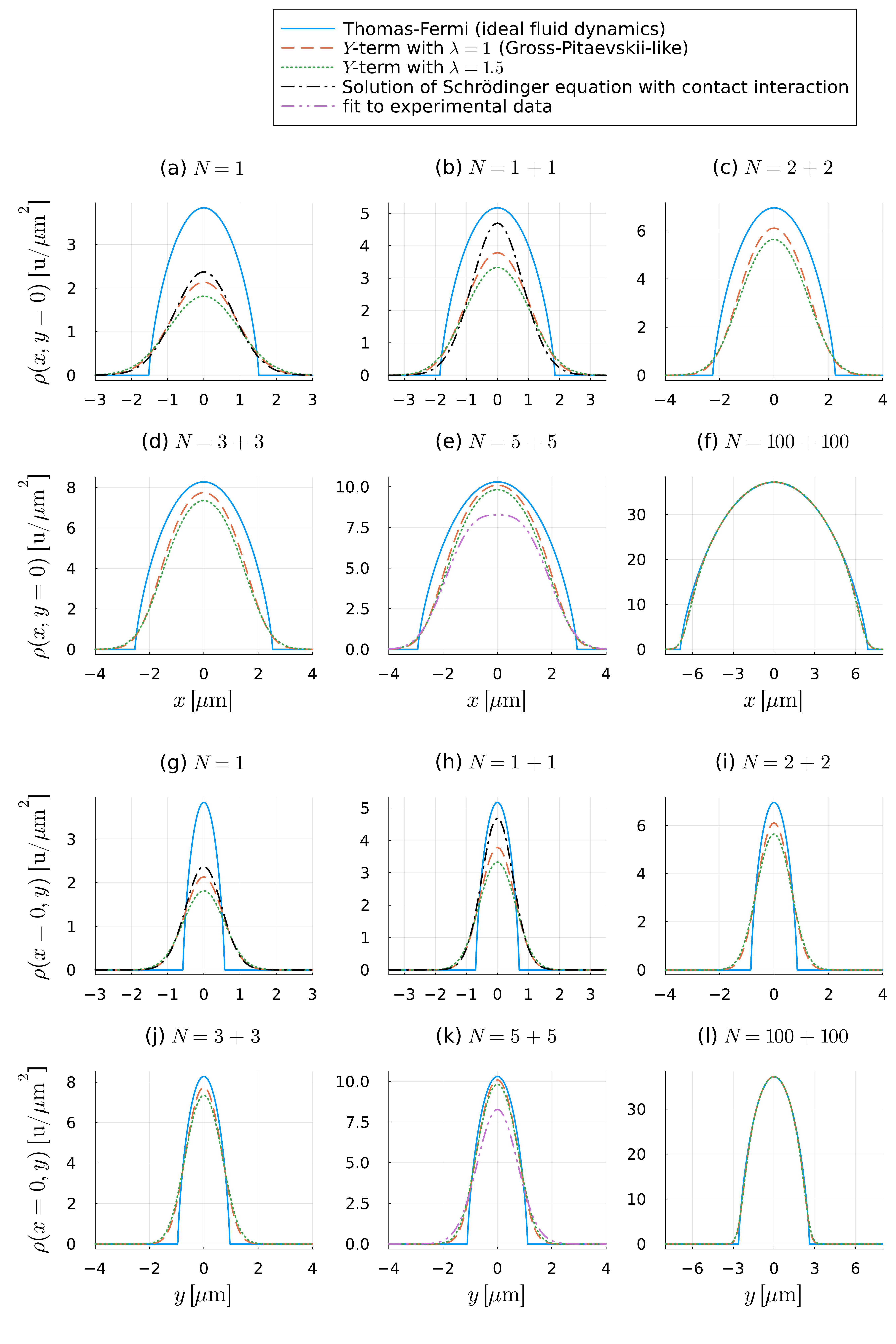}
    \caption{
        Comparison of cross sections of the predicted density profiles of a two-dimensional fluid in a harmonic trap for a Thomas-Fermi type solution (blue, solid) and the cases $\lambda=1$ (orange, dashed) and $\lambda=1.5$ (green, dotted) for the effective action ansatz presented in section \ref{sec:SecondOrderCorrections}. 
        The equation of state and trap frequencies are taken to be polytropic pressure with $\kappa = 2.335$, g = $7.06 \, \mathrm{u}^{1-\kappa} \mu \mathrm{m}^{-3\kappa - 1} \mathrm{ms}^{-2}$; $\omega_x = 1.280 \cdot 2\pi \, \mathrm{kHz}$ and $\omega_y = 3.384 \cdot 2\pi \, \mathrm{kHz}$, the same as in ref. \cite{Brandstetter:2023jsy}.
        The profiles are presented for a variety of particle numbers.
        Subfigures (a) to (f) show a cross-section along the plane defined by $y=0$, while subfigures (g) to (l) make the cut along $x=0$.
        For the single particle $N=1$ the ground state solution of the Schrödinger equation of a single (non-interacting) particle in a harmonic trap is shown for comparison (black, dash-dotted).
        For $N=1+1$ a ground state solution for a single pair of trapped particles with a contact interaction is taken from refs. \cite{Idziaszek_PhysRevA.74.022712, Liang_2008} (black, dash-dotted).
        For $N=5+5$ a phenomenological description of the density as a modified Gaussian from ref.\cite{Brandstetter:2023jsy} is shown for comparison (purple, dash-dot-dotted).
    \label{fig:DensityComparisonAlongX}
    }   
\end{figure*}

\begin{figure*}[t]
    \centering
    \includegraphics[width=0.9\textwidth]{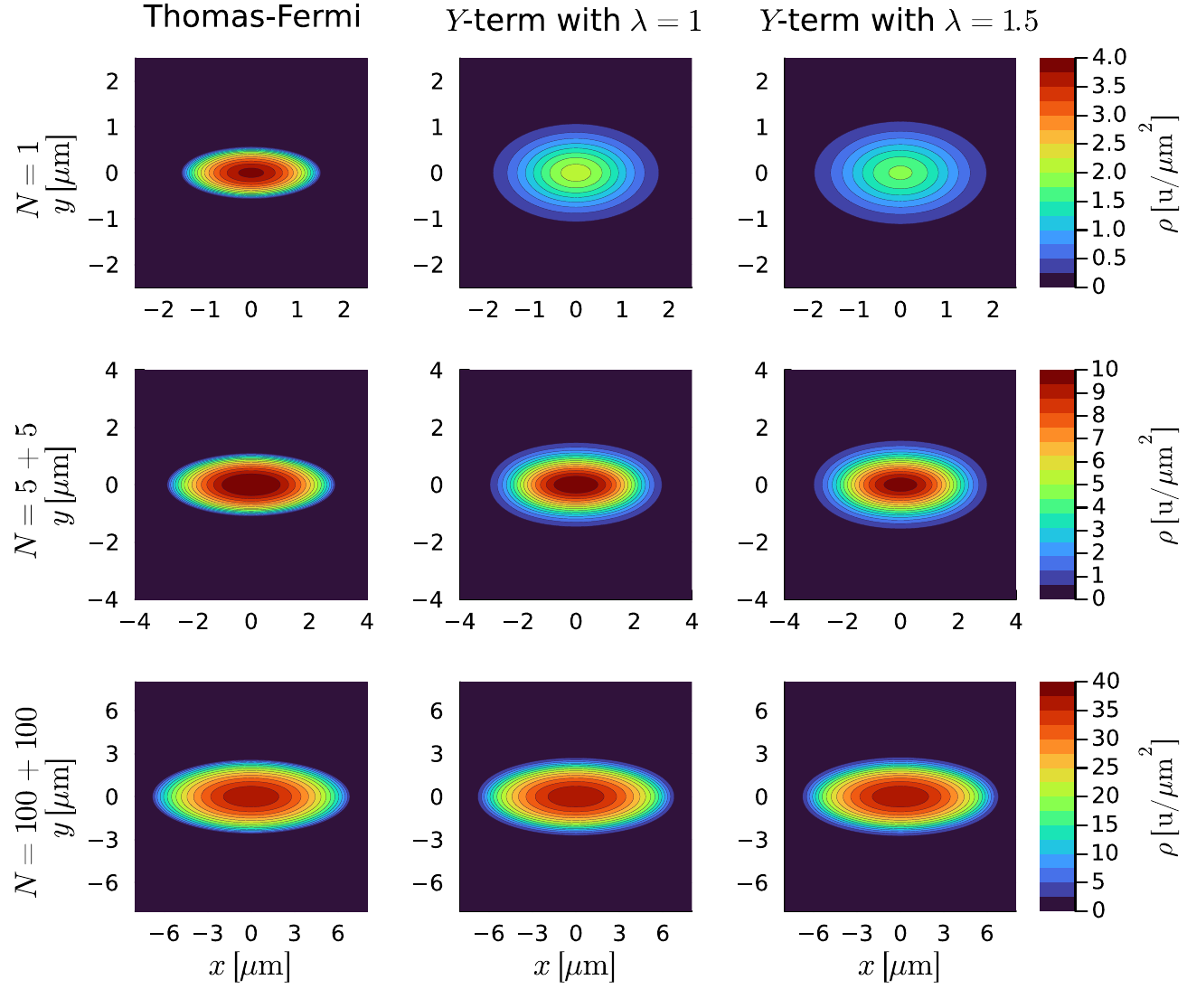}
    \caption{
        Comparison of the predicted density profiles of a two-dimensional fluid in a harmonic trap for a Thomas-Fermi type solution (left column) and the cases $\lambda=1$ (center column) and $\lambda=1.5$ (right column) for the effective action ansatz presented in section \ref{sec:SecondOrderCorrections}.
        The first row shows the predictions for a single particle, the second for ten particles, and the third for 200 particles.
        The equation of state and trap frequencies are taken to be polytropic pressure with $\kappa = 2.335$, g = $7.06 \, \mathrm{u}^{1-\kappa} \mu \mathrm{m}^{-3\kappa - 1} \mathrm{ms}^{-2}$; $\omega_x = 1.280 \cdot 2\pi \, \mathrm{kHz}$ and $\omega_y = 3.384 \cdot 2\pi \, \mathrm{kHz}$, the same as in ref. \cite{Brandstetter:2023jsy}.
    \label{fig:DensityComparison2d}
    }   
\end{figure*}

In Fig.~\ref{fig:DensityComparisonAlongX} we display trapped density profiles obtained from the numerical calculation. With reference to the effective action ansatz discussed in section \ref{sec:SpecificYTerm}, we show Thomas-Fermi distributions, corresponding to the solution with $\lambda=0$ (blue solid lines), as well as the result from Gross-Pitaevskii theory, with $\lambda=1$ (orange dashed lines). The results are presented for slices of the full two-dimensional densities (displayed instead in Fig.~\ref{fig:DensityComparison2d}), either $\rho(x,y=0)$ (two upper rows), or $\rho(x=0,y)$ (two lower rows). We show results for different particle numbers, or total mass in the system, where $N=1$ corresponds to the mass of an individual $^6$Li atom. Obviously, the main feature discerning the $\lambda=0$ from the $\lambda=1$ case is the smoothness of the distribution at the tails. For low particle numbers, even up to $N=10$, the effect is very pronounced, whereas as expected it becomes negligible in the many-body limit with $N=200$.

In addition, for the computation with $N=10$ atoms, we display as well the result of a fit to the initial density profile shown in ref.~\cite{Brandstetter:2023jsy} (double-dot-dashed line in purple). Remarkably, the tails of the experiment-based result, which are significantly broader than predicted by the Thomas-Fermi approximation, appear to be captured by Gross-Pitaevskii theory ($\lambda=1$). The description improves slightly for a higher value of $\lambda=1.5$ (also shown in all panels as green dotted lines),  which has the effect of broadening the tails a bit further.

Finally, for the trivial $N=1$ case (non-interacting particle), but also for $N=2$ (single interacting pair of fermions) we are able to display  exact analytical solutions of the Schrödinger equation (see e.g. \cite{Idziaszek_PhysRevA.74.022712, Liang_2008}). The discrepancy between the Gross-Pitaveskii curves and the exact ones arises from an interaction term which is included in the effective action description based on the order parameter. It would be interesting to have exact (or near-exact) solutions up to $N=6$ or higher, and see at which atom number a hydrostatic framework based on Gross-Pitaevskii theory becomes viable. If that happens with just 6 or 10 fermions, that may call for a re-thinking of the standard criteria for the application of Gross-Pitaveskii theory, which should in principle hold when fluctuations on top of the (classical) average field operator are negligible, which can not be the case when $1/\sqrt{N}\sim\mathcal{O}(1)$.

We move on now to the results of Fig.~\ref{fig:DensityComparison2d}, displaying trapped mass densities in the two-dimensional plane. This gives in particular a visualization of the impact of the second-order corrections on the edges of the clouds. We highlight the significant broadening for $N=10$ (panels in the middle of the figure), which depletes the anisotropy of the Thomas-Fermi distribution. Indeed, as noted in ref.~\cite{Brandstetter:2023jsy}, the expected scaling of the initial cloud anisotropy with the trap frequencies: $\langle x^2 \rangle / \langle y^2 \rangle = w_y^2/w_x^2$, is not observed in the experimental data, which favors a more isotropic initial condition. For $N=100$ (plots in the bottom), the impact of the second-order terms becomes negligible.

\subsection{Second order corrections to fluid dynamic evolution}

Beyond just the hydrostatics in the trap, we can also simulate the expansion of the fluid systems after the trap is removed.
For this we need to solve the continuity and momentum equations, which includes the second-order correction,
\begin{equation}
\begin{split}
    &\partial_t \rho + \boldsymbol{\nabla} \cdot (\rho \mathbf{v}) = 0 \, , \\
    &\rho (\partial_t + \mathbf{v} \cdot \boldsymbol{\nabla}) \mathbf{v} = -\boldsymbol{\nabla} p + \lambda \frac{\hbar^2}{2m^2} \rho \boldsymbol{\nabla} \left(\frac{\boldsymbol{\nabla}^2 \sqrt{\rho}}{\sqrt{\rho}} \right) \, ,
\end{split}
\end{equation}
with $\rho(t=0, \mathbf{x})$ taken from the hydrostatic solutions of section \ref{sec:hydrostatics}, and the fluid velocity set to zero everywhere initially $\mathbf{v}(t=0,\mathbf{x}) = 0$.
For the ideal fluid flow with Thomas-Fermi initial conditions, equivalent to $\lambda = 0$ in this ansatz, this can be reformulated into the ordinary differential equations \eqref{eq:scalingHydroEquations} and solved as before.
The solutions for the cases $\lambda=1$ and $\lambda=1.5$ use an equivalent evolution with the Hamiltonian in eq.~\eqref{eq:effectiveHamiltionianWithYTerm}, to evolve the order parameter $\varphi(t, \mathbf{x})$ as in the hydrostatic case.
The main difference here is the initial condition $\varphi(t=0, \mathbf{x}) = \sqrt{\rho(t=0, \mathbf{x})}$ and the evolution in real instead of imaginary time.
This is again done with a split step Fourier method.

\subsubsection{Real space expansion}

\begin{figure*}
    \centering
    \includegraphics[width=0.7\textwidth]{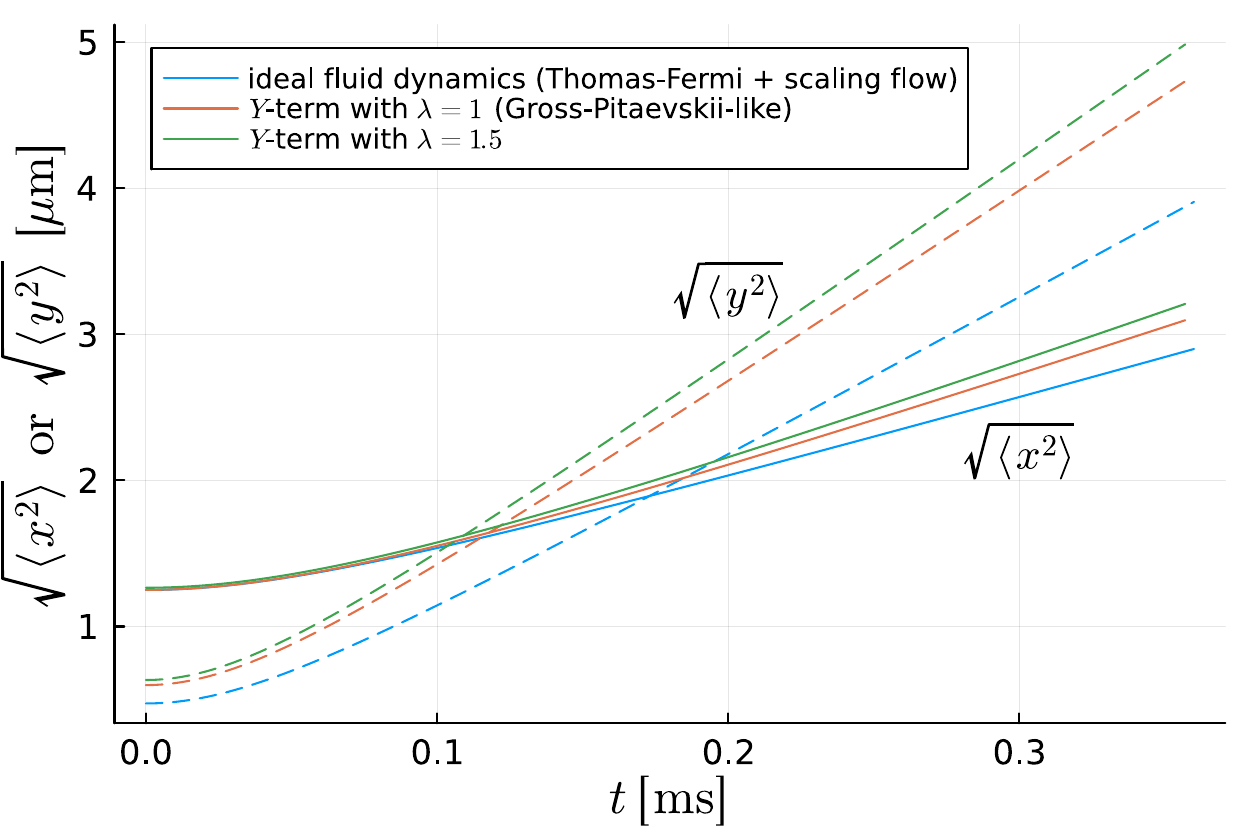}
    \caption{
        Expansion dynamics for a fluid released from equilibrium in a harmonic trap following ideal fluid dynamics (blue), the effective action ansatz of Sec. \ref{sec:SpecificYTerm} with $\lambda=1$, and with $\lambda=1.5$.
        Solid curves show the widths $\sqrt{\expval{x^2}}$ in $x$-direction, dashed curves the widths in $y$-direction.
        For the ideal fluid curves we solve eqs. \eqref{eq:scalingHydroEquations} with a fourth order accurate Rosenbrock method (Rodas4 solver in the Julia module DifferentialEquations).
        The other two curves were obtained using the Split-Step Fourier method with the Hamiltonian \eqref{eq:effectiveHamiltionianWithYTerm}.
    \label{fig:RealWidthEvolution}
    }   
\end{figure*}

The temporal evolution of the density-weighted root-mean-square value of the profile along the two axes, which we denote by
\begin{align}
    \sqrt{\langle x^2 \rangle}(t) &= \sqrt{\frac{\int d^2{\bf x}~ x^2 \rho({\bf x},t)}{\int d^2 {\bf x} ~\rho({\bf x},t)}} \, , \\
        \sqrt{\langle y^2 \rangle} (t)&= \sqrt{\frac{\int d^2{\bf x}~ y^2 \rho({\bf x},t)}{\int d^2{\bf x}~ \rho({\bf x},t)}} \, ,
\end{align}
 is shown in Fig.~\ref{fig:RealWidthEvolution}.
In general, for the same total mass and equation of state, a more squeezed initial condition should lead to an enhanced flow velocity, due to higher pressure gradients. We see, however, that the solutions for $\lambda=1$ and $\lambda=1.5$ (green and orange lines) evolve faster than the solution obtained for $\lambda=0$ (Thomas-Fermi approximation, depicted in blue lines), even though they start from a density profile that is more isotropic than the Thomas-Fermi one. This is, on one hand, good news, as the ideal hydrodynamic results shown in ref.~\cite{Brandstetter:2023jsy} appear to under-predict the velocity of the cloud evolution along both $x$ and $y$. On the other hand,  the inclusion of the $\lambda$ term significantly reduces the time at which the aspect ratio of the cloud becomes unity. This may lead to inconsistencies with the experiments, as the temporal evolution of the aspect ratio of the system is captured well by the ideal fluid expansion. More detailed investigations are needed to settle this.

\subsubsection{Considerations on the momentum space evolution}

We also briefly discuss the application of our results in momentum space, where experimental data is also available.  Indeed, experimentally it is not only possible to determine positions of the atoms at a given point in time, but also their momenta at early times during the expansion \cite{Subramanian_2023}.
This is done by a fast change in the magnetic field to bring the scattering length, $a$, close to zero.
Afterwards the particles are free streaming and their momenta get determined by the time of flight.
The momentum distribution $\dd{N}_i/\dd[3]{p}$ (where the index $i$ labels hyperfine spin) can be determined from repeated measurements.

While fluid dynamics by itself can not make predictions for quantities defined in momentum space, as discussed in ref.~\cite{Brandstetter:2023jsy}, one may attempt to associate certain combinations of fluid dynamic variables with specific moments of the measured momentum distribution. For this purpose, we investigate a quantity that corresponds to the volume integral of the momentum flux density,
\begin{equation}
    \mathcal{P}_{jk}(t) = \int d^Dx \mathscr{P}_{jk}(t,\mathbf{x}).
\label{eq:integratedMomentumFlux}
\end{equation}
This quantity is interesting because, for a non-interacting gas of (classical) particles with single particle phase space distribution $f_i(t,\mathbf{x},\mathbf{p})=dN_i/d^D x d^D p$, one has
\begin{equation}
    \mathscr{P}_{jk}(t, \mathbf{x}) = \sum_i \int d^D p \left\{\frac{p_j p_k}{m} f_i(t,\mathbf{x},\mathbf{p}) \right\} \, ,
\end{equation}
such that the integrated momentum flux,
\begin{equation}
    \mathcal{P}_{jk}(t) = \sum_i  \int d^D p \left\{ \frac{p_j p_k}{m} \frac{dN_i}{d^D p} \right\} \, ,
\end{equation}
becomes the second moment of the momentum distribution which can be determined experimentally.

Consider first the case of ideal fluid dynamics, where
\begin{equation}
    \mathcal{P}_{jk}(t) = \int d^D x \left\{ \rho(t,\mathbf{x}) v_j(t,\mathbf{x}) v_k(t,\mathbf{x}) + P(t,\mathbf{x}) \delta_{jk} \right\}.
\end{equation}
The integral can be easily determined for given fluid fields and equation of state. 
One should be cautious here, because the translation to momentum space through the phase space distribution function works only for non-interacting particles. 
 Experimentally one can change the interaction strength very quickly. What can happen during this transition? 
As a consequence of the conservation laws for particle number and momentum, the time derivatives of $\rho(t,\mathbf{x})$ and $\rho(t,\mathbf{x})v_j(t, \mathbf{x})$ must remain regular as functions of time. 
However, the pressure $P(t,\mathbf{x})$ can change during the quick ramp in interaction strength. 
In the simplest scenario it would change from the pressure associated to $\rho(t,\mathbf{x})$ in the interacting equation of state to the one associated with the same density $\rho(t,\mathbf{x})$ for a non-interacting equation of state. 
However, the ramp in interacting strength is a non-equilibrium process, and it is likely that a bulk viscous pressure is created, as well.

To avoid assumptions about the dynamics of the pressure term during the ramp one could study instead of $P_{jk}$ the trace-less tensor
\begin{equation}
    \hat{\mathcal{P}}_{jk}(t) = \mathcal{P}_{jk}(t) - \delta_{jk} \frac{1}{D}  \sum_{l=1}^D \mathcal{P}_{ll}(t).
\end{equation}
Based on eq.\ \eqref{eq:integratedMomentumFlux} and the decomposition in eq.\ \eqref{eq:decompositionMomentumFluxDensity} one finds 
\begin{equation}
\begin{split}
    \hat{\mathcal{P}}_{jk}(t) = \int d^D x\bigg\{ &\rho(t,\mathbf{x}) \Big{[} v_j(t, \mathbf{x}) v_k(t, \mathbf{x}) \\
    - \delta_{jk} \frac{1}{D} &\mathbf{v}(t,\mathbf{x})^2 \Big{]} + \pi_{jk}(t, \mathbf{x}) \bigg\}.
    \label{eq:tracelessIntegratedMomentumFlux}    
\end{split}
\end{equation}
Isotropic pressure and bulk viscous pressure terms have now been subtracted.

The first two terms on the right hand side of \eqref{eq:tracelessIntegratedMomentumFlux} are robust because they only depend on conserved densities $\rho(t,\mathbf{x})$ and $\rho(t,\mathbf{x}) v_j(t, \mathbf{x})$ that cannot change abruptly during a ramp in interaction strength. Based on this, ideal hydrodynamic predictions for the temporal evolution of the momentum anisotropy of the system, $\langle p_x^2 \rangle - \langle p_y^2 \rangle$ are shown in ref.~\cite{Brandstetter:2023jsy}, which turn out to be in overall good agreement with the experimental data.

What happens, then, to the second-order terms discussed here? Beyond the ideal fluid description, the symmetric and traceless shear stress tensor $\pi_{jk}(t, \mathbf{x})$ appears, and second-order corrections to this term were found in eq.\ \eqref{eq:secondOrderStressLambda}. It is likely that the shear stress terms (like the ones in eq.\ \eqref{eq:secondOrderStressLambda}) get modified by the ramp in interaction strength because they are not protected by conservation laws. Specifically, a ramp to zero interaction strength should destroy superfluidity and when the superfluid density $n$ is taken to zero, the shear stress terms in eq.\ \eqref{eq:secondOrderStressLambda} vanish. At least, this would preserve agreement with the experimental findings of ref.\ \cite{Brandstetter:2023jsy} that there is no momentum anisotropy at the initial time, right after the trap has been switched off.

In summary, without a more detailed understanding of the non-equilibrium processes happening during the short time interval in which the interaction strength is changed, it seems difficult to make robust statements about the impact of the second-order corrections on the integrated momentum flux, and consequently on the momentum anisotropy built up during the expansion, from fluid dynamic considerations alone.

\section{Conclusion}\label{sec:Conclusions}

We have discussed non-relativistic fluid dynamics at second order in the derivative expansion. For superfluids at zero temperature, we have emphasized contributions to the momentum flux density or stress tensor that involve second derivatives of the superfluid density. A well-known example is the quantum pressure contribution to the Gross-Pitaevskii description of bosons with weak repulsive contact interactions. We have argued that this represents a specific case of second-order contribution in a generic treatment based on an effective action ansatz. 

More specifically, the term of the form $Y(n_s) \boldsymbol{\nabla} n_s \boldsymbol{\nabla} n_s$ that appears in the effective action of eq.~\eqref{eq:QuantumEffectiveActionSuperfluidOrderParameter} allows us to modify the coefficient of the quantum pressure term.  In the context of density functional theory that describes static systems, a term of this kind has been discussed in the literature before and is known as von Weizsäcker term. Specifically for mesoscopic systems of strongly-interacting particles, there is \textit{a priori} no reason to assume that this new term should be neglected.

We have discussed, then, a hydrodynamic treatment of the mesoscopic Fermi gas analyzed in the experiments of ref.~\cite{Brandstetter:2023jsy}, encompassing both zeroth- and second-order (super)fluid dynamics.

Of particular interest is the modification of the Thomas-Fermi profile for the trapped density induced by the quantum corrections, with the second-order terms yielding much broader tails at the edges of the small systems, improving significantly the agreement with the experimental observations. In regards to the temporal evolution of the cloud of 10 strongly-interacting fermions in real space, quantum corrections tend to speed up the evolution of the cloud's rms sizes, also improving agreement with the experimental data, although comparisons with more experiments (possibly, time-resolved expansions for different atom numbers) are needed to draw any firm conclusions. Similarly, this will help shed  light on the impact of second-order corrections on the development of the momentum anisotropy in the expanding clouds, which we are not able to clarify at present. 

Indeed, we expect more data on the mesoscopic strongly-interacting Fermi gas to be available in the near future, as well as more results from more microscopic theoretical descriptions. This will allows us to learn more about the extent of applicability of a pure second-order hydrodynamic formulation for these small quantum systems. 
For further avenues of investigation, let us mention that the second-order coefficients investigated here could be calculated from microscopic theories, notably, in terms of response theory around a many-body ground or thermal state. In addition, these coefficients could be of relevance for other types of mesoscopic fluids, or boundary layers in larger systems, and warrant further study in general.

\section*{Acknowledgements}
The authors acknowledge funding by the Deutsche Forschungsgemeinschaft (DFG, German Research Foundation) – Project-ID 273811115 – SFB 1225 ISOQUANT, and under Germany's Excellence Strategy EXC2181/1-390900948 (the Heidelberg STRUCTURES Excellence Cluster). The authors acknowledge useful discussions with the participants of the EMMI Rapid Reaction Task Force \textit{``Deciphering many-body dynamics in mesoscopic quantum gases''}. The authors thank I.~Siovitz for useful recommendations on numerics, and S.~Brandstetter, T.~Enss, M.~Ga\l ka, C.~Heintze, M.~Holten, S.~Jochim, P.~Lunt, S.~Masciocchi, A.~Mazeliauskas, P.~M.~Preiss, I.~Selyuzhenkov, and K.~Subramanian for useful discussions and collaboration on related topics.

\appendix

\section{Improvement ambigutity for stresses at second order \label{app:Ambiguity}}

In this appendix, we present a more detailed discussion on terms arising in the fluid dynamic derivative expansion of stresses at second oder that are equivalent from the point of view of the equations of motion. This phenomenon is known as improvement ambiguity. We do this by working backwards from the terms appearing in the fluid dynamic equations, i.e. the combinations $\partial_j T_{ji}$ and $(\partial_i v_j) T_{ij} + \partial_i q_i$, and trying to invert these for the second order coefficients listed in Sec. \ref{sec:SecondOrderList}.
We include contributions in both number density (as in the main paper) and energy density, labeling the corresponding coefficients with the superscripts ${(n)}$ and ${(\varepsilon)}$ respectively.
Further, there are two terms which mix derivatives of number and energy density,
\begin{equation}
\begin{split}
    &\Pi^{(2,n\varepsilon)} = \alpha^{(n\varepsilon)} (\partial_i n) (\partial_i \varepsilon) \, , \\
    &\pi_{jk}^{(2,n\varepsilon)} = \beta^{(n\varepsilon)} (\partial_{<j} n) (\partial_{k>} \varepsilon) \, .
\end{split}
\end{equation}
Firstly, we decompose the derivative of the stress-energy tensor as it appears in the momentum equation into 
\begin{equation*}
\begin{split}
    \partial_j T_{ji}^{(2)} &=  F_1^{(n)} (\partial_i \partial_j \partial_j n) + F_2^{(n)} (\partial_i n) (\partial_j \partial_j n) \\
    &+ F_3^{(n)} (\partial_j n) (\partial_i \partial_j n) + F_4^{(n)} (\partial_i n) (\partial_j n) (\partial_j n) \\
    &+ F_1^{(\varepsilon)} (\partial_i \partial_j \partial_j \varepsilon) + F_2^{(\varepsilon)} (\partial_i \varepsilon) (\partial_j \partial_j \varepsilon) \\
    &+ F_3^{(\varepsilon)} (\partial_j \varepsilon) (\partial_i \partial_j \varepsilon) + F_4^{(\varepsilon)} (\partial_i \varepsilon) (\partial_j \varepsilon) (\partial_j \varepsilon) \\
    &+ G_1 (\partial_i \varepsilon) (\partial_j \partial_j n) + G_2 (\partial_i n) (\partial_j \partial_j \varepsilon) \\
    &+ G_3 (\partial_j \varepsilon) (\partial_i \partial_j n) + G_4 (\partial_j n) (\partial_i \partial_j \varepsilon) \\
    &+ G_5 (\partial_i \varepsilon) (\partial_j n) (\partial_j n) + G_6 (\partial_i n) (\partial_j \varepsilon) (\partial_j \varepsilon) \\
    &+ G_7 (\partial_i \varepsilon) (\partial_j \varepsilon) (\partial_j n) + G_8 (\partial_i n) (\partial_j n) (\partial_j \varepsilon) \\
    &+ H_1 (\partial_j v_j) (\partial_i \partial_k v_k) + H_2 (\partial_j v_j) (\partial_k \partial_k v_i) \\
    &+ H_3 (\partial_i v_j) (\partial_j \partial_k v_k) + H_4 (\partial_j v_i) (\partial_j \partial_k v_k) \\
    &+ H_5 (\partial_j v_k) (\partial_i \partial_j v_k) + H_6 (\partial_j v_k) (\partial_j \partial_k v_i) \\
    &+ H_7 (\partial_i v_j) (\partial_k \partial_k v_j) + H_8 (\partial_j v_i) (\partial_k \partial_k v_j) \\
    &+ H_9 (\partial_j v_k) (\partial_i \partial_k v_j) \\
    &+ I_1^{(n)} (\partial_i n) (\partial_j v_j) (\partial_k v_k) + I_2^{(n)} (\partial_j n) (\partial_i v_j) (\partial_k v_k) \\
    &+ I_3^{(n)} (\partial_j n) (\partial_j v_i) (\partial_k v_k) + I_4^{(n)} (\partial_i n) (\partial_j v_k) (\partial_j v_k) \\
    &+ I_5^{(n)} (\partial_j n) (\partial_k v_j) (\partial_k v_i) + I_6^{(n)} (\partial_i n) (\partial_j v_k) (\partial_k v_j) \\
    &+ I_7^{(n)} (\partial_j n) (\partial_j v_k) (\partial_k v_i) + I_8^{(n)} (\partial_j n) (\partial_j v_k) (\partial_i v_k) \\
    &+ I_9^{(n)} (\partial_j n) (\partial_k v_j) (\partial_i v_k) \\
\end{split}
\end{equation*}
\begin{equation}
\begin{split}
    &+ I_1^{(\varepsilon)} (\partial_i \varepsilon) (\partial_j v_j) (\partial_k v_k) + I_2^{(\varepsilon)} (\partial_j \varepsilon) (\partial_i v_j) (\partial_k v_k) \\
    &+ I_3^{(\varepsilon)} (\partial_j \varepsilon) (\partial_j v_i) (\partial_k v_k) + I_4^{(\varepsilon)} (\partial_i \varepsilon) (\partial_j v_k) (\partial_j v_k) \\
    &+ I_5^{(\varepsilon)} (\partial_j \varepsilon) (\partial_k v_j) (\partial_k v_i) + I_6^{(\varepsilon)} (\partial_i \varepsilon) (\partial_j v_k) (\partial_k v_j) \\
    &+ I_7^{(\varepsilon)} (\partial_j \varepsilon) (\partial_j v_k) (\partial_k v_i) + I_8^{(\varepsilon)} (\partial_j \varepsilon) (\partial_j v_k) (\partial_i v_k) \\
    &+ I_9^{(\varepsilon)} (\partial_j \varepsilon) (\partial_k v_j) (\partial_i v_k) \, .
\end{split}
\label{app:stressDerivativeFull}
\end{equation}
We split the terms by the types of derivatives they contain as every dependence on the thermodynamic variables $n$ and $\varepsilon$ can be absorbed into the coefficient functions.
We can now calculate the derivatives of all second order contributions (see eqs.\ \eqref{eq:secondOrderStressContributions} and \eqref{eq:secondOrderStressContributions2}) and compare the structure in derivatives to Eq. \eqref{app:stressDerivativeFull}.
For the terms containing only derivatives of the number density we find
\begin{equation}
\begin{split}
    &F_1^{(n)} = \alpha_1^{(n)} + \frac{D - 1}{D} \beta_1^{(n)} \, , \\
    &F_2^{(n)} = \partial_n \alpha_1^{(n)} - \frac{1}{D} \partial_n \beta_1^{(n)} + \beta_2^{(n)} \, , \\
    &F_3^{(n)} = 2 \alpha_2^{(n)} + \partial_n \beta_1^{(n)} + \frac{D - 2}{D} \beta_2^{(n)} \, , \\
    &F_4^{(n)} = \partial_n \alpha_2^{(n)} + \frac{D - 1}{D} \partial_n \beta_2^{(n)}\, .
\end{split}
\end{equation}
We do the same for the terms containing only derivatives of the energy density,
\begin{equation}
\begin{split}
    &F_1^{(\varepsilon)} = \alpha_1^{(\varepsilon)} + \frac{D - 1}{D} \beta_1^{(\varepsilon)} \, , \\
    &F_2^{(\varepsilon)} = \partial_\varepsilon \alpha_1^{(\varepsilon)} - \frac{1}{D} \partial_\varepsilon \beta_1^{(\varepsilon)} + \beta_2^{(\varepsilon)} \, , \\
    &F_3^{(\varepsilon)} = 2 \alpha_2^{(\varepsilon)} + \partial_\varepsilon \beta_1^{(\varepsilon)} + \frac{D - 2}{D} \beta_2^{(\varepsilon)} \, , \\
    &F_4^{(\varepsilon)} = \partial_\varepsilon \alpha_2^{(\varepsilon)} + \frac{D - 1}{D} \partial_\varepsilon \beta_2^{(\varepsilon)} \, , 
\end{split}
\end{equation}
the terms containing a combination of energy and number density derivatives,
\begin{equation}
\begin{split}
    &G_1 = \partial_\varepsilon \alpha_1^{(n)} - \frac{1}{D} \partial_\varepsilon \beta_1^{(n)} + \frac{1}{2} \beta^{(n\varepsilon)} \, , \\
    &G_2 = \partial_n \alpha_1^{(\varepsilon)} - \frac{1}{D} \partial_n \beta_1^{(\varepsilon)} + \frac{1}{2} \beta^{(n\varepsilon)} \, , \\
    &G_3 = \partial_\varepsilon \beta_1^{(n)} + \alpha^{(n\varepsilon)} + \frac{D - 2}{2D} \beta^{(n\varepsilon)} \, , \\
    &G_4 = \partial_n \beta_1^{(\varepsilon)} + \alpha^{(n\varepsilon)} + \frac{D - 2}{2D} \beta^{(n\varepsilon)} \, , \\
    &G_5 = \partial_\varepsilon \alpha_2^{(n)} - \frac{1}{D} \partial_\varepsilon \beta_2^{(n)} + \frac{1}{2} \partial_n \beta^{(n\varepsilon)} \, , \\
    &G_6 = \partial_n \alpha_2^{(\varepsilon)} - \frac{1}{D} \partial_n \beta_2^{(\varepsilon)} + \frac{1}{2} \partial_\varepsilon \beta^{(n\varepsilon)} \, , \\
    &G_7 = \partial_n \beta_2^{(\varepsilon)} + \partial_\varepsilon \alpha^{(n\varepsilon)} + \frac{D - 2}{2D} \partial_\varepsilon \beta^{(n\varepsilon)} \, , \\
    &G_8 = \partial_\varepsilon \beta_2^{(n)} + \partial_n \alpha^{(n\varepsilon)} + \frac{D - 2}{2D} \partial_n \beta^{(n\varepsilon)} \, , 
\end{split}
\end{equation}
the terms containing only derivatives of the fluid velocity,
\begin{equation}
\begin{split}
    &H_1 = 2 \alpha_3 + \frac{D - 4}{2D} \beta_3  \, , \\
    &H_2 = \frac{1}{2} \beta_3 \, , \\
    &H_3 = \frac{1}{2} \beta_3 + \frac{1}{2} \beta_6  \, , \\
    &H_4 = \frac{1}{2} \beta_3 + \beta_4  \, , \\
    &H_5 = 2 \alpha_4 - \frac{2}{D} \beta_4  + \frac{D - 2}{D} \beta_5 \, , \\
    &H_6 = \beta_4 + \frac{1}{2} \beta_6 \, , \\
    &H_7 = \beta_5 \, , \\
    &H_8 = \frac{1}{2} \beta_6 \, , \\
    &H_9 = 2 \alpha_5 + \frac{D - 4}{2D} \beta_6 \, , 
\end{split}
\end{equation}
the terms combining derivatives of fluid velocity and number density,
\begin{equation}
\begin{split}
    &I_1^{(n)} = \partial_n \left( \alpha_3 - \frac{1}{D} \beta_3 \right) \, , \\
    &I_2^{(n)} = \frac{1}{2} \partial_n \beta_3 \\
    &I_3^{(n)} = \frac{1}{2} \partial_n \beta_3 \\
    &I_4^{(n)} = \partial_n \left( \alpha_4 - \frac{1}{D} \beta_4 - \frac{1}{D} \beta_5 \right) \, , \\
    &I_5^{(n)} = \partial_n \beta_4 \, , \\
    &I_6^{(n)} = \partial_n \left( \alpha_5 - \frac{1}{D} \beta_6 \right) \, , \\
    &I_7^{(n)} = \frac{1}{2} \partial_n \beta_6 \, , \\
    &I_8^{(n)} = \partial_n \beta_5 \, , \\
    &I_9^{(n)} = \frac{1}{2} \partial_n \beta_6\, , 
\end{split}
\end{equation}
and finally the terms combining derivatives of fluid velocity and energy density,
\begin{equation*}
\begin{split}
    &I_1^{(\varepsilon)} = \partial_\varepsilon \left( \alpha_3 - \frac{1}{D} \beta_3 \right) \, , \\
    &I_2^{(\varepsilon)} = \frac{1}{2} \partial_\varepsilon \beta_3 \, , \\
    &I_3^{(\varepsilon)} = \frac{1}{2} \partial_\varepsilon \beta_3 \, , \\
    &I_4^{(\varepsilon)} = \partial_\varepsilon \left( \alpha_4 - \frac{1}{D} \beta_4 - \frac{1}{D} \beta_5 \right) \, , \\
\end{split}
\end{equation*}
\begin{equation}
\begin{split}
    &I_5^{(\varepsilon)} = \partial_\varepsilon \beta_4 \, , \\
    &I_6^{(\varepsilon)} = \partial_\varepsilon \left( \alpha_5 - \frac{1}{D} \beta_6 \right) \, , \\
    &I_7^{(\varepsilon)} = \frac{1}{2} \partial_\varepsilon \beta_6 \, , \\
    &I_8^{(\varepsilon)} = \partial_\varepsilon \beta_5 \, , \\
    &I_9^{(\varepsilon)} = \frac{1}{2} \partial_\varepsilon \beta_6 \, .
\end{split}
\end{equation}
These sets of equations can be split into two blocks which are independent of one another.
The first block contains all contributions from the second order corrections that are independent of velocity, i.e. the equations involving $F^{(n)}$, $F^{(\varepsilon)}$, and $G$.
These can be combined to obtain several contraint equations for the coefficient functions which are automatically fulfilled by the assumed structure of the stress-energy tensor,
\begin{equation}
\begin{split}
    &F_4^{(n)} = \frac{1}{2} \partial_n (F_2^{(n)} + F_3^{(n)} - \partial_n F_1^{(n)}) \, , \\
    &F_4^{(\varepsilon)} = \frac{1}{2} \partial_\varepsilon ( F_2^{(\varepsilon)} + F_3^{(\varepsilon)} - \partial_\varepsilon F_1^{(\varepsilon)}) \, , \\
    &G_5 = \partial_n G_1 + \frac{1}{2} \partial_\varepsilon (F_3^{(n)} - \partial_n F_1^{(n)} - F_2^{(n)}) \, , \\
    &G_6 = \partial_\varepsilon G_2 + \frac{1}{2} \partial_n (F_3^{(\varepsilon)} - \partial_\varepsilon F_1^{(\varepsilon)} - F_2^{(\varepsilon)}) \, , \\
    &G_8 = \partial_n G_3 + \partial_\varepsilon F_2^{(n)} - \partial_n \partial_\varepsilon F_1^{(n)} \, , \\
    &G_7 = \partial_\varepsilon G_4 + \partial_n F_2^{(\varepsilon)} - \partial_\varepsilon \partial_n F_1^{(\varepsilon)} \, , \\
    &G_1 - G_2 + G_3 - G_4 = \partial_\varepsilon F_1^{(n)} - \partial_n F_1^{(\varepsilon)} \, .
\end{split}
\end{equation}
This means that there are 10 coefficients $\alpha_1^{(n)}, \alpha_2^{(n)}, \beta_1^{(n)}, \beta_2^{(n)}, \alpha_1^{(\varepsilon)}, \alpha_2^{(\varepsilon)}, \beta_1^{(\varepsilon)}, \beta_2^{(\varepsilon)}, \alpha^{(n\varepsilon)}$ and~$\beta^{(n\varepsilon)}$ and only 9 leftover equations, i.e. the system of equations is underdetermined.
We can choose one of these coefficients freely and still end up with any given second order hydrodynamic equations.
Without loss of generality we will use $\beta_1^{(n)}$ as this degree of freedom in the following.
There is one more freedom of choice which comes in via the connection between the $n$- and $\varepsilon$-coefficients since one sector can only constrain the derivatives of the other.
The connecting equation is
\begin{equation}
    \partial_n \beta_1^{(\varepsilon)} - \partial_\varepsilon \beta_1^{(n)} = G_1 - G_2 + \partial_n F_1^{(\varepsilon)} - \partial_\varepsilon F_1^{(n)} \, .
\end{equation}
This means that $\beta_1^{(\varepsilon)}$ is only determined by a choice of $\beta_1^{(n)}$ up to an arbitrary added function in energy density which will be called $\Delta \beta_1^{(\varepsilon)}$ in the following.

The second block of coefficient functions contains all contributions from the velocity dependent second order corrections, i.e. the remaining equations involving $H$, $I^{(n)}$, and $I^{(\varepsilon)}$.
These can be combined to obtain
\begin{equation}
\begin{split}
    &H_6 = H_3 + H_4 - 2H_2 \, , \\
    &H_8 = H_3 - H_2 \, , \\
    &I_1^{(n)} = \frac{1}{2} \partial_n (H_1 - H_2) \, , \\
    &I_2^{(n)} = I_3^{(n)} = \partial_n H_2 \, , \\
    &I_4^{(n)} = \frac{1}{2} \partial_n (H_5 - H_7) \, , \\
    &I_5^{(n)} = \partial_n (H_4 - H_2) \, , \\
    &I_6^{(n)} = \frac{1}{2} (H_9 - H_6) \, , \\
    &I_7^{(n)} = I_9^{(n)} = \partial_n (H_3 - H_2) \, , \\
    &I_8^{(n)} = \partial_n H_7 \, ,
\end{split}
\end{equation}
and constraints of the same structure also exist for the $I_k^{(\varepsilon)}$.
In total, 20 of the 27 equations reduce to constraints on the coefficient functions, the remaining 7 can be used to uniquely identify $\alpha_3, \alpha_4, \alpha_5, \beta_3, \beta_4, \beta_5$ and $\beta_6$.

The other way in which the heat flux and stress-energy tensor are involved in the fluid dynamic equations is as part of the energy equation.
We again split the second order terms by their derivate structure,
\begin{equation}
\begin{split}
    (\partial_i v_j) &T_{ij}^{(2)} + \partial_i q_i^{(2)} = J (\partial_i \partial_i \partial_j v_j) \\
    &+ K_1^{(n)} (\partial_i n) (\partial_j \partial_j v_i) + K_2^{(n)} (\partial_i n) (\partial_i \partial_j v_j) \\
    &+ K_3^{(n)} (\partial_i v_i) (\partial_j \partial_j n) + K_4^{(n)} (\partial_i v_j) (\partial_i \partial_j n) \\
    &+ K_1^{(\varepsilon)} (\partial_i \varepsilon) (\partial_j \partial_j v_i) + K_2^{(\varepsilon)} (\partial_i \varepsilon) (\partial_i \partial_j v_j) \\
    &+ K_3^{(\varepsilon)} (\partial_i v_i) (\partial_j \partial_j \varepsilon) + K_4^{(\varepsilon)} (\partial_i v_j) (\partial_i \partial_j \varepsilon) \\
    &+ L_1^{(n)} (\partial_i v_i) (\partial_j n) (\partial_j n) + L_2^{(n)} (\partial_i v_j) (\partial_i n) (\partial_j n) \\
    &+ L_1^{(\varepsilon)} (\partial_i v_i) (\partial_j \varepsilon) (\partial_j \varepsilon) + L_2^{(\varepsilon)} (\partial_i v_j) (\partial_i \varepsilon) (\partial_j \varepsilon) \\
    &+ M_1 (\partial_i v_i) (\partial_j n) (\partial_j \varepsilon) + M_2 (\partial_i v_j) (\partial_i n) (\partial_j \varepsilon) \\
    &+ M_3 (\partial_i v_j) (\partial_i \varepsilon) (\partial_j n) + \mathcal{O}(v^3) \, .
    \label{eq:ambiguityHeatFluxPart}
\end{split}
\end{equation}
The terms of third order in velocity can only come from the velocity dependent second order contributions to the stress-energy tensor.
Since the coefficients contained therein are already fixed, the resulting equations will not add any new information and we will not consider them in the following.
For the terms in equation \eqref{eq:ambiguityHeatFluxPart} that are of first order in the velocities we can identify
\begin{equation*}
\begin{split}
    &J = \gamma_4 + \gamma_5 \, , \\
    &K_1^{(n)} = \gamma_1^{(n)} + \partial_n \gamma_4 \, , \\
    &K_2^{(n)} = \gamma_2^{(n)} + \gamma_3^{(n)} + \partial_n \gamma_5 \, , \\
    &K_3^{(n)} = \alpha_1^{(n)} - \frac{1}{D} \beta_1^{(n)} + \gamma_3^{(n)} \, , \\
    &K_4^{(n)} = \beta_1^{(n)} + \gamma_1^{(n)} + \gamma_2^{(n)} \, , \\
\end{split}
\end{equation*}
\begin{equation}
\begin{split}
    &K_1^{(\varepsilon)} = \gamma_1^{(\varepsilon)} + \partial_\varepsilon \gamma_4 \, , \\
    &K_2^{(\varepsilon)} = \gamma_2^{(\varepsilon)} + \gamma_3^{(\varepsilon)} + \partial_\varepsilon \gamma_5 \, , \\
    &K_3^{(\varepsilon)} = \alpha_1^{(\varepsilon)} - \frac{1}{D} \beta_1^{(\varepsilon)} + \gamma_3^{(\varepsilon)} \, , \\
    &K_4^{(\varepsilon)} = \beta_1^{(\varepsilon)} + \gamma_1^{(\varepsilon)} + \gamma_2^{(\varepsilon)} \, , \\
    &L_1^{(n)} = \alpha_2^{(n)} - \frac{1}{D} \beta_2^{(n)} + \partial_n \gamma_3^{(n)} \, , \\
    &L_2^{(n)} = \beta_2^{(n)} + \partial_n \gamma_1^{(n)} + \partial_n \gamma_2^{(n)} \, , \\
    &L_1^{(\varepsilon)} = \alpha_2^{(\varepsilon)} - \frac{1}{D} \beta_2^{(\varepsilon)} + \partial_\varepsilon \gamma_3^{(\varepsilon)} \, , \\
    &L_2^{(\varepsilon)} = \beta_2^{(\varepsilon)} + \partial_\varepsilon \gamma_1^{(\varepsilon)} + \partial_\varepsilon \gamma_2^{(\varepsilon)} \, , \\
    &M_1 = \alpha^{(n\varepsilon)} - \frac{1}{D} \beta^{(n\varepsilon)} + \partial_\varepsilon \gamma_3^{(n)} + \partial_n \gamma_3^{(\varepsilon)} \, , \\
    &M_2 = \frac{1}{2} \beta^{(n\varepsilon)} + \partial_n \gamma_1^{(\varepsilon)} + \partial_\varepsilon \gamma_2^{(n)} \, , \\
    &M_3 = \frac{1}{2} \beta^{(n\varepsilon)} + \partial_\varepsilon \gamma_1^{(n)} + \partial_n \gamma_2^{(\varepsilon)} \, ,
\end{split}
\end{equation}
which connect the non-velocity-dependent parts of the second order stress-energy tensor to the 8 coefficients in the second order heat flux ($\gamma_1^{(n)}, \gamma_2^{(n)}, \gamma_3^{(n)}, \gamma_1^{(\varepsilon)}, \gamma_2^{(\varepsilon)}, \gamma_3^{(\varepsilon)}, \gamma_4$ and $\gamma_5$).
Of the 16 equations, we find that 9 can be reduced to relations between the coefficient functions,
\begin{equation}
\begin{split}
    K_1^{(n)} + K_2^{(n)} {}&{} - K_3^{(n)} - K_4^{(n)} = \partial_n J - F_1^{(n)} \, , \\
    K_1^{(\varepsilon)} + K_2^{(\varepsilon)} {}&{} - K_3^{(\varepsilon)} - K_4^{(\varepsilon)} = \partial_\varepsilon J - F_1^{(\varepsilon)} \, , \\
    L_1^{(n)} = \partial_n K_3^{(n)} {}&{} + \frac{1}{2} (F_3^{(n)} - F_2^{(n)} - \partial_n F_1^{(n)}) \, , \\
    L_2^{(n)} = \partial_n K_4^{(n)} {}&{} + F_2^{(n)} - \partial_n F_1^{(n)} \, , \\
    L_1^{(\varepsilon)} = \partial_\varepsilon K_3^{(\varepsilon)} {}&{} + \frac{1}{2} (F_3^{(\varepsilon)} - F_2^{(\varepsilon)} - \partial_\varepsilon F_1^{(\varepsilon)}) \, , \\
    L_2^{(\varepsilon)} = \partial_\varepsilon K_4^{(\varepsilon)} {}&{} + F_2^{(\varepsilon)} - \partial_\varepsilon F_1^{(\varepsilon)} \, , \\
    M_1 = \partial_n K_3^{(\varepsilon)} {}&{} + \partial_\varepsilon K_3^{(n)} - \frac{1}{2} (\partial_n F_1^{(\varepsilon)} + \partial_\varepsilon F_1^{(n)}) \\
    {}&{} + \frac{1}{2} (G_3 + G_4 - G_1 - G_2) \, , \\
    M_2 + M_3 = {}&{} \partial_n K_4^{(\varepsilon)} + \partial_\varepsilon K_4^{(n)} + G_1 + G_2 \\
    {}&{} - \partial_n F_1^{(\varepsilon)} - \partial_\varepsilon F_1^{(n)} \, , \\
    M_2 - M_3 = {}&{} G_1 - G_2 + \partial_n (K_1^{(\varepsilon)} + K_3^{(\varepsilon)} - K_2^{(\varepsilon)}) \\
    {}&{} - \partial_\varepsilon (K_1^{(n)} + K_3^{(n)} - K_2^{(n)}) \, ,
\end{split}
\end{equation}
such that from the heat flux section we gain an additional coefficient function that we can choose freely which without loss of generality we will use $\gamma_5$ for.
That means that for given hydrodynamic behavior there are in total two functions in both number and energy density and one in just either number or energy density that one can choose freely among the second order coefficients.
For our choice of freely chosen $\beta_1^{(n)}$, $\Delta \beta_1^{(\varepsilon)}$ and $\gamma_5$, the other coefficients are uniquely determined by the coefficient functions to be
\begin{equation}
\begin{split}
    \beta_1^{(\varepsilon)} = {}&{} \int^n_{n_0} \dd{n'} (\partial_\varepsilon \beta_1^{(n)} + G_1 - G_2 + \partial_n F_1^{(\varepsilon)} - \partial_\varepsilon F_1^{(n)}) \, , \\
    {}&{} + \Delta \beta_1^{(\varepsilon)} \, , \\
    \alpha_1^{(n)} = {}&{} \frac{1 - D}{D} \beta_1^{(n)} + F_1^{(n)} \, , \\
    \alpha_2^{(n)} = {}&{} \frac{1 - D}{D} \partial_n \beta_1^{(n)} + \frac{1}{2} \bigg{(} F_3^{(n)} - \frac{D - 2}{D} F_2^{(n)} \, , \\
    {}&{} - \frac{D - 2}{D} \partial_n F_1^{(n)} \bigg{)} \, , \\
    \beta_2^{(n)} = {}&{} \partial_n \beta_1^{(n)} + F_2^{(n)} - \partial_n F_1^{(n)} \, , \\
    \alpha_1^{(\varepsilon)} = {}&{} \frac{1 - D}{D} \beta_1^{(\varepsilon)} + F_1^{(\varepsilon)} \, , \\
    \alpha_2^{(\varepsilon)} = {}&{} \frac{1 - D}{D} \partial_\varepsilon \beta_1^{(\varepsilon)} + \frac{1}{2} \bigg{(} F_3^{(\varepsilon)} - \frac{D - 2}{D} F_2^{(\varepsilon)} \, , \\
    {}&{} - \frac{D - 2}{D} \partial_\varepsilon F_1^{(\varepsilon)} \bigg{)} \, , \\
    \beta_2^{(\varepsilon)} = {}&{} \partial_\varepsilon \beta_1^{(\varepsilon)} + F_2^{(\varepsilon)} - \partial_\varepsilon F_1^{(\varepsilon)} \, , 
\end{split}
\end{equation}
\begin{equation}
\begin{split}
    \alpha^{(n\varepsilon)} = {}&{} \frac{2 - 2D}{D} \partial_\varepsilon \beta_1^{(n)} + G_3 - \frac{D - 2}{D} G_1 \\
    {}&{} + \frac{D - 2}{D} \partial_\varepsilon F_1^{(n)} \, , \\
    \beta^{(n\varepsilon)} = {}&{} 2 \partial_\varepsilon \beta_1^{(n)} + 2G_1 - 2\partial_\varepsilon F_1^{(n)} \, , 
\end{split}
\end{equation}
\begin{equation}
\begin{split}
    &\alpha_3 = \frac{1}{2} H_1 - \frac{D - 4}{2D} H_2 \, , \\
    &\alpha_4 = \frac{1}{2} H_5 + \frac{1}{D} (H_4 - H_2) - \frac{D - 2}{2D} H_7 \, , \\
    &\alpha_5 = \frac{1}{2} H_9 - \frac{D - 4}{2D} (H_3 - H_2) \, , \\
    &\beta_3 = 2 H_2 \, , \\
    &\beta_4 = H_4 - H_2 \, , \\
    &\beta_5 = H_7 \\
    &\beta_6 = 2 (H_3 - H_2) \, , 
\end{split}
\end{equation}
\begin{equation}
\begin{split}
    &\gamma_1^{(n)} = \partial_n \gamma_5 + K_1^{(n)} - \partial_n J \, , \\
    &\gamma_2^{(n)} = -\beta_1^{(n)} - \partial_n \gamma_5 + K_2^{(n)} - K_3^{(n)} + F_1^{(n)} \, , \\
    &\gamma_3^{(n)} = \beta_1^{(n)} + K_3^{(n)} - F_1^{(n)} \, , \\
    &\gamma_1^{(\varepsilon)} = \partial_\varepsilon \gamma_5 + K_1^{(\varepsilon)} - \partial_\varepsilon J \, , \\
    &\gamma_2^{(\varepsilon)} = -\beta_1^{(\varepsilon)} - \partial_\varepsilon \gamma_5 + K_2^{(\varepsilon)} - K_3^{(\varepsilon)} + F_1^{(\varepsilon)} \, , \\
    &\gamma_3^{(\varepsilon)} = \beta_1^{(\varepsilon)} + K_3^{(\varepsilon)} - F_1^{(\varepsilon)} \, , \\
    &\gamma_4 = J - \gamma_5 \, . 
\end{split}
\end{equation}

\section{Scaling solutions using Lagrangian coordinates \label{app:LagrangeCoords}}

In this appendix we discuss the expansion after release from a (possibly anisotropic) harmonic trap. Within an ideal fluid description this problem can be almost solved analytically, at least strongly reduced in complexity from partial differential equations to ordinary differential equations. This is known in the literature, see for example ref.\ \cite{PitaevskiiStringari2016}. We present here a discussion of this phenomenon in terms of Lagrangian coordinates.

\subsection{Lagrangian coordinate description}

Lagrangian coordinates are coordinates that move with the fluid. This is based on the map from positions $\mathbf{R}$ at some reference time $t_0$ to the position $\mathbf{r}$ at time $t$, denoted by $\mathbf{r}(t, \mathbf{R})$. This map always exists and is determined by the fluid velocity $\mathbf{v}(t,\mathbf{x})$ through the integral
\begin{equation}
  \mathbf{r}(t, \mathbf{R}) = \mathbf{R} + \int_{t_0}^t dt^\prime \, \mathbf{v}(t^\prime, \mathbf{r}(t^\prime, \mathbf{R})). 
\end{equation}
For us it is convenient to take $t_0$ to be the time where the trap is switched off. Obviously the fluid velocity is in turn given by
\begin{equation}
  \mathbf{v}(t,\mathbf{r}(t, \mathbf{R})) = \frac{\partial}{\partial t} \mathbf{r}(t, \mathbf{R}).
\end{equation}
We shall also assume that $\mathbf{r}(t, \mathbf{R})$ is an invertible map, with inverse $\mathbf{R}(t, \mathbf{r})$. This is not always guaranteed, and breaks down when streams of matter cross. In the following we will deliberately use either $\mathbf{x}$ or $\mathbf{R}$ or $\mathbf{r}$ as coordinates with the understanding that they are related by $\mathbf{x} = \mathbf{r}(t,\mathbf{R})$. 

We also need the Jacobi matrix
\begin{equation}
  J_{jk}(t,\mathbf{R}) = \frac{\partial}{\partial R_j} r_k(t, \mathbf{R}),
\end{equation}
and its inverse given by
\begin{equation}
  I_{kj}(t,\mathbf{r}) = \frac{\partial}{\partial r_k} R_j(t, \mathbf{r}),
\end{equation}
evaluated at coinciding points.

The great benefit of Lagrangian coordinates is that the Galilei covariant derivative becomes just a partial derivative at fixed $\mathbf{R}$,
\begin{equation}
  \left( \frac{\partial}{\partial t} + v_j(t,\mathbf{x}) \frac{\partial}{\partial x_j} \right) \rho(t, \mathbf{x}) = \frac{\partial}{\partial t} \rho(t, \mathbf{r}(t, \mathbf{R})){\big |}_{\mathbf{R}}. 
\end{equation}
On the other side, the digergence of the fluid velocity becomes now
\begin{equation}
\begin{split}
  & \boldsymbol{\nabla} \cdot \mathbf{v}(t,\mathbf{x}) = \frac{\partial}{\partial x_k} \frac{\partial}{\partial t} r_k(t,\mathbf{R}) \\ & = \left[\frac{\partial }{\partial r_k}R_j(t, \mathbf{r}) \right] \frac{\partial}{\partial t}\frac{\partial}{\partial R_j} r_k(t,\mathbf{R}) = I_{kj}(t,\mathbf{r}) \frac{\partial}{\partial t} J_{jk}(t, \mathbf{R}).
\end{split}
\end{equation}
Using Jacobis formula one can rewrite this as
\begin{equation}
  \boldsymbol{\nabla} \cdot \mathbf{v}(t,\mathbf{x}) = \frac{1}{J(t,\mathbf{R})} \frac{\partial}{\partial t} J(t,\mathbf{R}),
\end{equation}
where $J(t,\mathbf{R}) = \text{det}(J_{jk}(t,\mathbf{R}))$ is the Jacobi determinant. 

We are now ready to formulate the continuity equation is Lagrangian coordinates,
\begin{equation}
  \begin{split}
    \left( \frac{\partial}{\partial t} + v_j(t,\mathbf{x}) \frac{\partial}{\partial x_j} \right) \rho(t, \mathbf{x}) + \rho(t, \mathbf{x}) \boldsymbol{\nabla} \mathbf{v}(t,\mathbf{x}) = & \\
    \frac{\partial}{\partial t} \rho(t,  \mathbf{R}) + \frac{\rho(t, \mathbf{R})}{J(t,\mathbf{R})} \frac{\partial}{\partial t} J(t,\mathbf{R}) = & 0.
  \end{split}
\end{equation}
This implies simply that $\rho(t, \mathbf{R}) J(t,\mathbf{R})$ is constant in time. A similar relation holds for other conserved densities, for example entropy density $s(t, \mathbf{r}(t, \mathbf{R}))$ for an ideal fluid. Ratios of such densities, such as the entropy per particle $s/n$ for an ideal fluid are independent of time in Lagrangian coordinates.

Let us now discuss the Euler equation, which reads in Euler coordinates
\begin{equation}
  \begin{split}
    &\left( \frac{\partial}{\partial t} + v_j(t,\mathbf{x}) \frac{\partial}{\partial x_j} \right) v_k(t,\mathbf{x}) + \frac{1}{m n(t, \mathbf{x})} \frac{\partial}{\partial x_k} P(t,\mathbf{x}) =  \\
    &\left( \frac{\partial}{\partial t} + v_j(t,\mathbf{x}) \frac{\partial}{\partial x_j} \right) v_k(t,\mathbf{x}) + \frac{1}{m} \frac{\partial}{\partial x_k} \mu(t,\mathbf{x}) \\
    &+ \frac{s(t,\mathbf{x})}{m n(t,\mathbf{x})} \frac{\partial}{\partial x_k} T(t,\mathbf{x}) = \, 0 \, .
  \end{split}
\end{equation}
In the second equation we used the differential $dp = n d \mu + s dT$. 
In Lagrangian coordinates we find accordingly
\begin{equation}
\begin{split}
    & \frac{\partial^2}{\partial t^2} r_k(t, \mathbf{R}) + \frac{1}{m} \left[\frac{\partial}{\partial x_k} R_j(t, \mathbf{x}) \right] \\ & \times \left[ \frac{\partial}{\partial R_j} \mu(t,\mathbf{R}) + \frac{s(t,\mathbf{R})}{n(t,\mathbf{R})} \frac{\partial}{\partial R_j} T(t,\mathbf{R}) \right] = 0.    
\end{split}
\end{equation}
The second term involves the inverse Jacobi matrix. 
By multiplying both sides with the Jacobi matrix $J_{jk}(t,\mathbf{R})$ we obtain
\begin{equation}
\begin{split}
  & m J_{jk}(t, \mathbf{R})\frac{\partial^2}{\partial t^2} r_k(t, \mathbf{R}) + \frac{\partial}{\partial R_j} \mu(t,\mathbf{R}) \\ &+ \frac{s(t,\mathbf{R})}{n(t,\mathbf{R})} \frac{\partial}{\partial R_j} T(t,\mathbf{R}) = 0.
\end{split}\label{eq:EulerEquationLagrangian}
\end{equation}
Together with the conservation laws for mass and entropy densities and a thermodynamic equation of state, this provides a full set of evolution equations for an ideal fluid.

\subsection{Solution by ansatz in Lagrangian coordinates}
Consider again eq.\ \eqref{eq:EulerEquationLagrangian}. In the initial state at $t=t_0$ the chemical potential is quadratic in $\mathbf{R}$ and the temperature is initially constant. Accordingly the second term in the above relation is linear in $\mathbf{R}$ and the third term vanishes. Also $\mathbf{r}(t, \mathbf{R})$ is initially linear in $\mathbf{R}$ and accordingly $J_{jk}(t, \mathbf{R})$ is initially constant. It is an interesting question to ask whether such a simple behavior can persist at later times.

To make this more precise, let us make the following ansatz
\begin{equation}
  r_k(t,\mathbf{R}) = R_j J_{jk}(t),
  \label{eq:ansatz}
\end{equation}
with a time-dependent but spatially constant Jacobi matrix $J_{jk}(t)$. With this ansatz also the Jacobi determinant $J(t)$ is of course constant in space. 

The fluid velocity is obtained as
\begin{equation}
  v_k(t,\mathbf{R}) = R_j \dot J_{jk}(t),
\end{equation}
and its rotational part is
\begin{equation}
\begin{split}
    & \epsilon_{jkl} \frac{\partial}{\partial x_k} v_l(t,\mathbf{x}) = \epsilon_{jkl} I_{km}(t) \frac{\partial}{\partial R_m} v_l(t,\mathbf{R}) \\ & = \epsilon_{jkl} I_{km}(t)  \dot J_{ml}(t).  
\end{split}
\end{equation}
For an irrotational flow $J_{jk}(t)$ is diagonal. For completeness we also give the shear tensor
\begin{equation}
  \begin{split}
    \sigma_{ij}(t) = & \frac{1}{2} \frac{\partial}{\partial x_i} v_j + \frac{1}{2} \frac{\partial}{\partial x_j} v_i - \frac{1}{D} \delta_{ij} \boldsymbol{\nabla} \mathbf{v} \\
    = & \frac{1}{2} I_{ik}(t) \dot J_{kj}(t) + \frac{1}{2} I_{jk}(t) \dot J_{ki}(t) - \frac{1}{D} J(t)^{-1} \dot J(t).
  \end{split}
\end{equation}
Noteably it is independent of spatial position and vanishes in the isotropic case where $J_{jk}(t) = J(t)^{1/D} \delta_{jk}$.

For the mass and entropy densities the ansatz yields within ideal fluid dynamics the simple scaling solutions (using $J(t_0)=1$),
\begin{equation}
  \rho(t, \mathbf{R})  = \frac{\rho(t_0, \mathbf{R}) }{J(t)}, \quad\quad\quad s(t, \mathbf{R})  = \frac{s(t_0, \mathbf{R}) }{J(t)}.
\end{equation}
This implies in particular for the ratio
\begin{equation}
  (s/n)(t,\mathbf{R}) = (s/n)(t_0,\mathbf{R}).
\end{equation}
We now assume an equation of state of the form $\varepsilon(s,n) = \mu_0 n + g(s/n) n^\kappa/(\kappa-1)$. Using $d\varepsilon = \mu dn + T ds$ and $\varepsilon+p = \mu n + s T$ this implies $p = g(s/n) n^\kappa$ which generalizes the ansatz in eq.\ \eqref{eq:polytropicPresEOS} to non-vanishing entropy per particle $s/n$. With $\mu_0$ we denote a constant that corresponds to the chemical potential in vacuum and can be taken as $\mu_0= - E_b/2 = - 1/(2ma^2)$ with $E_b$ the binding energy of the shallow dimer.

For this equation of state one finds also
\begin{equation}
  (T/(\mu-\mu_0))(t,\mathbf{R}) = (T/(\mu-\mu_0))(t_0,\mathbf{R}),
\end{equation}
and
\begin{equation}
\begin{split}
\mu(t,\mathbf{R})-\mu_0 = & \frac{\mu(t_0,\mathbf{R})-\mu_0}{J(t)^{\kappa-1}}, \\
T(t,\mathbf{R}) = & \frac{T(t_0,\mathbf{R})}{J(t)^{\kappa-1}},
\end{split}
\end{equation}
In particular when the initial temperature is independent of $\mathbf{R}$, as we assume, this remains to be the case also for $t>t_0$. This makes clear that the ansatz \eqref{eq:ansatz} leads to a consistent solution of \eqref{eq:EulerEquationLagrangian}.

Specifically, the hydrostatic Thomas-Fermi profile \eqref{eq:ThomasFermiChemicalPotential} implies
\begin{equation}
  \frac{\partial}{\partial R_j} \mu(t,\mathbf{R}) = - m \omega^2_{jk} R_k \frac{1}{J(t)^{\kappa-1}} .
\end{equation}
This is linear in $\mathbf{R}$ also for $t>t_0$. We also have
\begin{equation}
  \frac{\partial^2}{\partial t^2} r_k(t, \mathbf{R}) = R_l \ddot J_{lk}(t),
\end{equation}
which is also linear. 

The Euler equation \eqref{eq:EulerEquationLagrangian} is then equivalent to a set of ordinary differential equations for the components of the matrix $J_{jk}(t)$,
\begin{equation}
  J(t)^{\kappa-1} J_{jk}(t) \ddot J_{lk}(t) = \omega^2_{jl}.
\end{equation}
For a cloud that is initially at rest the initial values at $t=t_0$ are given by
\begin{equation}
  J_{jk}(t_0) = \delta_{jk}, \quad\quad\quad \dot J_{jk}(t_0) = 0.
\label{eq:eomJ_app}
\end{equation}
In that case $J_{jk}(t)$ will remain diagonal also at later times in the frame where $\omega^2_{jk}$ is diagonal.

Interestingly eq.\ \eqref{eq:eomJ} provides a simple set of equations that fully determine the flow profile and it is independent of the function $g(s/n)$ in the equation of state, as well as the initial temperature and particle number. It only depends on the exponent $\kappa$ and the shape of the harmonic trap!

\bibliographystyle{apsrev4-2}
\bibliography{biblio}

\end{document}